\documentclass[twocolumn]{aastex62}
\pdfoutput=1 
\usepackage{amsmath,amstext}
\usepackage[T1]{fontenc}
\usepackage{apjfonts} 
\usepackage[figure,figure*]{hypcap}
\usepackage{natbib}
\usepackage[disable,colorinlistoftodos]{todonotes}

\usepackage{soul, CJK}

\let\Oldtodo\todo
\renewcommand{\todo}[1]{\Oldtodo[inline]{#1}}

\shorttitle{The Zwicky Transient Facility Scheduler}
\shortauthors{Bellm et al.}

\begin{document}
\begin{CJK*}{UTF8}{bkai}

\title{The Zwicky Transient Facility: Surveys and Scheduler}

\author[0000-0001-8018-5348]{Eric C. Bellm}
\affiliation{DIRAC Institute, Department of Astronomy, University of Washington, 3910 15th Avenue NE, Seattle, WA 98195, USA}
\affiliation{ecbellm@uw.edu}

\author[0000-0001-5390-8563]{Shrinivas R. Kulkarni}
\affiliation{Division of Physics, Mathematics, and Astronomy, California Institute of Technology, Pasadena, CA 91125, USA}

\author{Tom Barlow}
\affiliation{Division of Physics, Mathematics, and Astronomy, California Institute of Technology, Pasadena, CA 91125, USA}

\author[0000-0002-9435-2167]{Ulrich Feindt}
\affiliation{The Oskar Klein Centre, Department of Physics, Stockholm University, AlbaNova, SE-106 91 Stockholm, Sweden}  

\author[0000-0002-3168-0139]{Matthew J. Graham}
\affiliation{Division of Physics, Mathematics, and Astronomy, California Institute of Technology, Pasadena, CA 91125, USA}

\author[0000-0002-4163-4996]{Ariel Goobar}
\affiliation{The Oskar Klein Centre, Department of Physics, Stockholm University, AlbaNova, SE-106 91 Stockholm, Sweden}

\author[0000-0002-6540-1484]{Thomas Kupfer}
\affiliation{Kavli Institute for Theoretical Physics, University of California, Santa Barbara, CA 93106, USA}
\affiliation{Department of Physics, University of California, Santa Barbara, CA 93106, USA}
\affiliation{Division of Physics, Mathematics, and Astronomy, California Institute of Technology, Pasadena, CA 91125, USA}

\author[0000-0001-8771-7554]{Chow-Choong Ngeow}
\affiliation{Institute of Astronomy, National Central University, 32001, Taiwan}

\author[0000-0002-3389-0586]{Peter Nugent}
\affiliation{Computational Science Department, Lawrence Berkeley National Laboratory, 1 Cyclotron Road, MS 50B-4206, Berkeley, CA 94720, USA}
\affiliation{Department of Astronomy, University of California, Berkeley, CA 94720-3411, USA}

\author{Eran Ofek}
\affiliation{Benoziyo Center for Astrophysics and the Helen Kimmel Center for Planetary Science, Weizmann Institute of Science, 76100 Rehovot, Israel}

\author{Thomas A. Prince}
\affiliation{Division of Physics, Mathematics, and Astronomy, California Institute of Technology, Pasadena, CA 91125, USA}

\author{Reed Riddle}
\affiliation{Caltech Optical Observatories, California Institute of Technology, Pasadena, CA 91125, USA} 

\author{Richard Walters}
\affiliation{Caltech Optical Observatories, California Institute of Technology, Pasadena, CA 91125, USA}

\author[0000-0002-4838-7676]{Quan-Zhi Ye (葉泉志)}
\affiliation{IPAC, California Institute of Technology, 1200 E. California Blvd, Pasadena, CA 91125, USA}
\affiliation{Division of Physics, Mathematics, and Astronomy, California Institute of Technology, Pasadena, CA 91125, USA}

\begin{abstract}

We present a novel algorithm for scheduling the observations of time-domain imaging surveys.
Our Integer Linear Programming approach optimizes an observing plan for an entire night by assigning targets to temporal blocks, enabling strict control of the number of exposures obtained per field and minimizing filter changes. 
A subsequent optimization step minimizes slew times between each observation.
Our optimization metric self-consistently weights contributions from time-varying airmass, seeing, and sky brightness to maximize the transient discovery rate.
We describe the implementation of this algorithm on the surveys of the Zwicky Transient Facility and present its on-sky performance.

\end{abstract}

\listoftodos

\section{Introduction}

Astronomers observe with telescopes costing millions or even billions of dollars that have finite useful lifetimes.
They must accordingly decide how best to sequence observations in order to maximize the scientific output of their facilities.
Despite its ubiquity, however, this scheduling problem remains challenging for both theoretical and practical reasons.
With hundreds to many thousands of observations to obtain in a night or a season, the potential number of observing sequences is combinatorically large: one thousand exposures have $\sim10^{2567}$ possible orderings.
The need to slew between targets couples distinct observations together, so maintaining efficiency requires scheduling many targets at once.
Observing conditions on the ground change rapidly, and requests for Target of Opportunity (TOO) observations can upend a carefully tuned schedule in an instant.
The quality of a potential observation may vary with time (e.g., with seeing, airmass, or moon phase), and many facilities must impose complex pointing or instrument constraints.
Time-domain surveys may require complex observing sequences that make future observations dependent on when past observations occurred, which is further complicated by weather losses and other downtime.
And finally it is often both difficult and impolitic to be quantitatively precise about how to measure scientific output.

Accordingly, direct, manual sequencing of observations by humans remains common at both ground- and space-based facilities.
Skilled operations staff can perform complex heuristic tradeoffs to obtain observations that are ``good enough'' while meeting the necessary constraints.
Satisficing in this manner may in some cases be the most efficient use of the human resources available, given the difficulties of developing more automated approaches.
However, manual scheduling presents significant drawbacks.
It is labor-intensive, requiring constant staffing throughout the operation of the project.
Any change to the schedule requires manual intervention, removing the ability to respond dynamically to changing observing conditions, weather losses, and TOOs.
Typically the scheduling process is not optimizing in any quantitative sense, limiting clarity about its effectiveness.
Because schedule generation is labor-intensive, it is difficult to compare different observing plans.
It is difficult to reproduce manual scheduling outcomes, which can inhibit studies of survey detection rates and efficiencies.
Finally, manual scheduling provides limited transparency to the user community about how their resources are being allocated.

``Greedy'' algorithms provide a convenient entry point into automated scheduling and are widely used in astronomy.
Before each observation, such algorithms compute a updated metric or score for each possible target, select the target with the current highest value, observe it, and then repeat the process.
Greedy optimizers are straightforward to implement, can easily handle changes to observing plans and conditions, provide traceable quantitative optimization, and can be run in an automated fashion.
The \texttt{Astroplan} package \citep{Morris:18:Astroplan} implements one such greedy scheduler.  
It is designed for human observers and implements a range of observational constraints.

However, it is widely recognized that local optimizers such as the greedy algorithm cannot deliver global optimization.  
For ground-based imaging surveys this problem manifests itself in a tendency to observe higher elevation fields as they rise instead of fields transiting at lower elevation at the same time.  
Some authors have recommended additional weighting schemes that account for the time until a target sets in order to penalize this behavior \citep[e.g.,][]{Denny:04:Scheduler, Denny:06:Scheduler}, but this is not the same as looking ahead to determine the optimal time to observe a given field.
Lookahead is especially valuable for time-domain surveys, which must understand how repeated observations of a target can be scheduled within the planning interval.

Several projects have implemented more sophisticated schedulers \citep[see][for a review]{Solar:16:SchedulingModel}.
Of particular note is the scheduling approach of the Las Cumbres Observatory (LCO)\footnote{\url{https://lco.global/}}. 
LCO operates a global network of replicated 0.6\,m, 1.0\,m, and 2.0\,m telescopes with identical imagers and spectrographs.
Rather than manually directing their observations to a specific telescope, LCO users make requests to the entire network, leaving to the scheduler the task of determining which facility to use to observe a target.
\citet{Lampoudi:15:LCOGTScheduler} describe the scheduling algorithm, which uses Integer Linear Programming 
(ILP\footnote{ILP problems have variables which take only discrete integer values, linear objective functions, and linear constraints.  Mixed ILP problems include some non-discrete variables.})
techniques to assign requested observations to telescopes subject to any observability or cadence constraints.
The LCO scheduler optimizes the assignments in order to maximize the total number of observations obtained, weighted by the priority assigned to them by the Time Allocation Committee (TAC).
Notably, the scheduler's ability to rapidly re-solve the entire network within minutes allows rapid TOO observations to be integrated into the scheduling process without disruption, as each new optimization run starts \textit{de novo}, integrating any new targets that have arrived in the meantime.

\citet{Solar:16:SchedulingModel} presents a similar Mixed Integer Linear Programming solution in the scheduler for ALMA.
The ALMA scheduler discretizes time into scheduling blocks and assigns observations to them in order to maximize TAC-assigned scientific priorities, program completion, and telescope utilization.
However, this scheme makes scheduling some types of observations relevant for time-domain followup challenging \citep{Alexander:17:AlmaTDA}.

Finally, \citet{Naghib:18:FeatureBasedScheduler} casts the scheduling problem of the Large Synoptic Survey Telescope (LSST) as a memoryless Markov Decision Process, using hand-designed features to reduce the dimensionality of the state space and optimizing the feature weights with a simple throughput-based objective function.

In this work we consider the specific scheduling problem of a single-telescope ground-based wide-field imaging survey.
We are focused on its application to the Zwicky Transient Facility \citep[ZTF;][]{tmp_Bellm:18:ZTFOverview, tmp_Graham:18:ZTFScience} project, which imposes some specific requirements (\S \ref{sec:requirements}), but our formalism is relevant for other time-domain surveys, such as those conducted with the Large Synoptic Survey Telescope \citep[LSST;][]{ivezic2008lsst}, the Dark Energy Camera \citep[DECam;][]{Flaugher:15:DECam}, and Hyper Suprime-Cam \citep[HSC;][]{Miyazaki:18:HSC}.
Minor modifications would enable its use by multi-telescope surveys such as the Asteroid Terrestrial-impact Last Alert System \citep[ATLAS;][]{Tonry:18:ATLAS}, PanSTARRS \citep{Kaiser:10:PanSTARRSSurvey},  the All-Sky Automated Survey for Supernovae \citep[ASAS-SN;][]{Shappee:14:ASASSN}, and BlackGEM \citep{Bloemen:16:BlackGEM}.

Simply stated, the scheduling problem to be solved is to determine which fields to observe in what order, with a goal of maximizing an objective function (\S \ref{sec:metric}; here, a proxy for the transient discovery rate) while achieving the desired temporal spacing of observations (``cadence'').
Optimizing the survey schedule provides a greater quantity of high-quality data, increasing the scientific output of the survey.
During the development of the ZTF survey camera and observing system \citep{tmp_Dekany:18:ZTFObservingSystem}, the engineering team devoted substantial effort to developing percent-scale improvements in throughput and efficiency.
Preserving these gains requires similar attention to the operation of the survey itself.
Our approach provides a self-consistent means of scheduling an entire night of ZTF observations.

In this paper we outline the scheduling approach used by ZTF and its application to the surveys undertaken during the early operations period.
In \S \ref{sec:requirements} we outline the requirements we used to guide our scheduler development.
\S \ref{sec:metric} describes the scalar survey speed metric we optimize for and discuss its applicability to other surveys and optimization approaches.
\S  \ref{sec:algorithm} presents the integer linear programming formalism we use to optimize ZTF observations for an entire night.
In \S \ref{sec:implementation} we describe the practical implementation the algorithm in the ZTF scheduler.
\S \ref{sec:surveys} details the surveys executed by ZTF in its first year of on-sky operations.
\S \ref{sec:performance} assesses the performance of the scheduler.
We conclude in \S \ref{sec:conclusions}.

\section{ZTF Requirements} \label{sec:requirements}

The requirements for the ZTF scheduler grew from the experience of the Palomar Transient Factory \citep[PTF;][]{Law:09:PTFOverview} and Intermediate Palomar Transient Factory (iPTF) surveys.  
PTF used a greedy scheduler. 
Its objective function is described in \citet{Law:09:PTFOverview}; it includes ad-hoc weightings for sun altitude, sky brightness excess due to the moon, moon phase, telescope and dome slews, airmass, and a cadence term.
In practice it proved unpredictable and prone to long slews. 
Operations staff frequently applied manual weights to ensure fields were observed.

For iPTF, a single member of the operations staff scheduled each night manually.  
This procedure reduced the number of long slews and (in conjunction with other technical improvements) increased the overall number of exposures taken.

During ZTF development, we began evaluating scheduling approaches in conjunction with other efforts at maximizing survey efficiency \citep{tmp_Dekany:18:ZTFObservingSystem}.
The vastly improved readout speed of ZTF (8\,sec) relative to PTF/iPTF (40\,sec) made limiting scheduling overheads a higher priority.
Additionally, the large number of simultaneous survey programs (\S \ref{sec:surveys}), some of them public, also necessitated the ability to simulate and test survey plans.
		
Specific requirements imposed on the ZTF scheduler included:
\begin{itemize}
    \item Select pointings from a fixed field grid \citep[see][]{tmp_Masci:18:ZTFDS}
    \item Operate in both simulation mode and on-sky using the same scheduling code
    \item Conduct several surveys (\S \ref{sec:surveys}), maintaining strict independence of their observations and balancing observing time between programs
    \item Provide interfaces for conducting Target of Opportunity observations and monitoring scheduler status
    \item Recover appropriately from interruptions and weather losses
    \item Maximize an observing efficiency metric and prioritize cadence control.
\end{itemize}

\section{Optimization Metric} \label{sec:metric}

The ZTF scheduler attempts to maximize (\S \ref{sec:algorithm}) the total number of exposures taken per night, weighted by the spatial volume probed by each, and subject to the constraints imposed by program balance and cadence (\S \ref{sec:surveys}).
If the observing cadences are well-chosen, maximizing this quantity will maximize the transient discovery rate.
\citet{Bellm:16:Cadences} explores the relationship between the chosen observing cadences, a survey's volumetric and areal survey rates, and the transient detection rate.

Neglecting cosmological effects, the volume $V_{\rm lim}$ probed by a given exposure is proportional to the cube of the limiting distance $d_{\rm lim}$ a transient of fiducial absolute magnitude $M$ can be detected given the limiting magnitude $m_{\rm lim}$: $V_{\rm lim} \propto d_{\rm lim}^3$, where $d = 10^{0.2 (m_{\rm lim} - M + 5)}$\,pc \citep[cf.][]{Bellm:16:Cadences}.
The volumetric weighting per exposure is thus 
\begin{equation}
V = 10^{0.6 (m_{\rm lim} - 21)} \label{eqn:V}
\end{equation}
where we have absorbed constant factors and normalized to a convenient limiting magnitude for ZTF.

This weighting combines in a self-consistent way many factors that are intuitively relevant for assessing whether an image is ``good'': the limiting magnitude depends on the filter, seeing, airmass, and sky brightness.
We use a model (\S \ref{sec:implementation}) to predict the variation in limiting magnitude and hence our metric as a function of these time-varying inputs.
Accordingly, our optimization will naturally select exposures near zenith and away from the moon; but by combining them in a single scalar the optimization can coherently trade these factors against one another as they change through the night.

Our metric deliberately does not contain factors that account for relative scientific priority or cadence.
These concerns have no general quantitative relationship to our objective function or each other\footnote{One could imagine a global model for the information contributed by a potential new observation given the past history of observations at that location, but we expect that this approach would require computationally expensive lightcurve modeling within the optimization loop and likely be limited to a single class of objects such as SN Ia.}.
Instead, we use the structure of the optimization algorithm (\S \ref{sec:algorithm}) to impose these constraints.

Our optimization algorithm (\S \ref{sec:algorithm}) maximizes the summed metric over an entire night.
In cases where a greedy algorithm is more convenient, it is simple to define an instantaneous volumetric survey speed
\begin{equation}
\dot{V} \propto 10^{0.6 m_{\rm lim}} / (t_{\rm exp} + t_{\rm OH})
\label{eqn:Vdot}
\end{equation}
that normalizes the volume probed in an exposure by the time required to obtain it, a sum of the exposure time $t_{\rm exp}$ and any readout or slew overheads $t_{\rm OH}$.

Other optimization metrics will be more appropriate for surveys
such as LSST \citep{ivezic2008lsst} that are concerned with coadded depth in addition to transient discovery; these may easily be substituted in our algorithm (\S \ref{sec:algorithm}).
For instance, \citet{Tonry:11:ATLAS}
suggests a 
weighting factor derived from information theory with a metric proportional to $10^{0.8 m_{\rm lim}}$.

\section{Algorithm} \label{sec:algorithm}

The ZTF scheduling process begins with a set of Observing Programs.
These are defined by their footprint on the sky (a discrete set of fields, which may be larger than the set observable in one night or even one lunation);
the fraction of the telescope observing time they are allocated;
the number of nights between successive revisits to this field for this program\footnote{Observing programs are not allowed to couple their observing sequences to the observing history of other programs;
each is completely independent.} (i.e., the ``inter-night gap,'' such as a 1-day or 3-day cadence);
and the number of visits and filter set for observations requested within a night (e.g., two nightly visits, one in $g$-band and one in $r$-band).

At the beginning of the night, each Observing Program provides a list of fields that are visible long enough to obtain the requested observations and have not been observed within that program's inter-night gap.
The resulting input to the scheduling algorithm are ``Request Sets:'' each request set is a ZTF field along with the number of observations requested per filter, the exposure time per observation, and appropriate Observing Program metadata.
For example, one Request Set might be for field 123 with three $g$-band and three $r$-band exposures tonight, all 30\,sec exposures: six Requests in total.
The scheduler also uses the past observation history, the fraction of time allocated to each Observing Program, and the length of the night to determine the number of allowed requests per Observing Program.

The scheduling algorithm then determines which request sets are observed, at what time to schedule the constituent observations, and how to arrange the slews and filter changes to maximize efficiency.
We use Integer Linear Programming (ILP) techniques to solve this problem; our notation and approach is inspired by that of \citet{Lampoudi:15:LCOGTScheduler}, but there are significant differences which we discuss in \S \ref{sec:other_ilp}.

\subsection{Parameters}

We construct the observing schedule by dividing the night into a set of temporal blocks $T$.
This discretization is necessary to make scheduling the entire night computationally tractable: rather than determining an exact sequence of hundreds or thousands of exposures, we must merely assign the observations to 15--25 blocks.
The block structure also provides a useful means of applying cadence constraints and minimizing filter changes (\S \ref{sec:constraints}).

The length of the block $T_{\rm block}$ is set to the minimum desired separation between exposures.
For ZTF we set the time block size to 30\,minutes, sufficient to identify the motion of main-belt asteroids with ZTF's moderate image quality ($\sim$2$^{\prime\prime}$ FWHM).

The set of available filters in the camera is $F$.
The set of Request Sets from all Observing Programs $P$ is $R$.
For each Request Set we use Equation \ref{eqn:V} to calculate the volumetric weighting factor $V_{rtf}$ for an observation of that field at time block $t \in T$ for filter $f \in F$.
The weight of an observation thus changes through the night: image quality, atmospheric transmission, and sky brightness change as fields rise and set, and the sky brightness also changes with the motion of the sun and moon.
We approximate the weight factor as constant within any single time block.
Filter changes only occur at the block boundaries.

\subsection{Decision Variables}

We solve for binary decision variables:

\begin{itemize}
    \item $Y_{rtf} = 1$ if Request Set $r \in R$ has an observation scheduled at time block $t \in T$ using filter $f \in F$, and 0 otherwise
\end{itemize}

We also define resultant variables used to apply constraints (\S \ref{sec:constraints}):

\begin{itemize}
    \item $Y_{tf} = 1$ if observations in time block $t \in T$ are conducted using filter $f \in F$, and 0 otherwise
    \item $Y_{s} = 1$ if the filter changes between time blocks $s \in T$ and $s+1 \in T$, and 0 otherwise
\end{itemize}

\subsection{Objective} 

The optimizer maximizes an objective function which sums the volume-weighted (Equation \ref{eqn:V}) number of exposures scheduled through the night.
Because of how we constrain the number of exposures in a temporal block (\S \ref{sec:constraints}), we also penalize for exposures lost due to filter changes.
The objective function is thus

\begin{equation}
    \mathrm{max} \left(\left(\sum_{r \in R} \sum_{t \in T} \sum_{f \in F} V_{rtf} Y_{rtf}\right) - \left(\frac{t_{\rm filt}}{t_{\rm exp} + t_{\rm OH}} w \sum_{t \in T} Y_s\right) \right)
    \label{eqn:objective}
\end{equation}
where $t_{\rm filt}$ is the time required to change filters and $w$ is a weight factor ($\approx \mathrm{max}(V_{rtf})$) accounting for the value of each lost exposure.

\subsection{Constraints} \label{sec:constraints}

Each scheduled Request Set should have exactly the requested number of observations per filter $n_{rf}$:
\begin{equation}
    \sum_{t \in T} Y_{rtf} = n_{rf}, \forall f \in F~\forall r \in R
    \label{eqn:constraint_nreqs}
\end{equation}

Only one filter should be used within a given time block:
\begin{equation}
    \sum_{f \in F} Y_{tf} = 1, \forall t \in T
    \label{eqn:constraint_onefilter}
\end{equation}

The time required to execute the observations assigned to a block should be less than or equal to the length of the block:
\begin{equation}
    \sum_{r \in R} \sum_{f \in F} Y_{rtf} \left(t_{{\rm exp,}r} + t_{\rm OH}\right) \leq T_{\rm block}, \forall t \in T
    \label{eqn:constraint_nperslot}
\end{equation}
where we have here allowed for variable exposure times per request set.
Because we have not yet \textit{sequenced} the observations in a block (see \S \ref{sec:tsp}), we don't know the exact slew times required and so use a fiducial value of 9\,sec (corresponding to the limit imposed by CCD readout) for the overhead time.
In practice this means a few more exposures may be scheduled in a block than can practically be observed.

Finally we apply a constraint to limit the number of scheduled requests per Observing Program to enforce the desired balance between programs:
\begin{equation}
    \sum_{r \in R, p^\prime = p} \sum_{t \in T} \sum_{f \in F} Y_{rtf} \leq n_{p^\prime}, \forall p \in P
    \label{equation:constraint_balance}
\end{equation}
where the number of allowed exposures $n_p$ for a given Observing Program $p$ is determined each night from the fractional observing time assigned to the program, the length of the requested exposures, and the past observing history.

\subsection{Sequencing Exposures within a block} \label{sec:tsp}

The solution to the ILP program is a list of observations assigned to each time block in the night.
We use a second optimization process to sequence observations efficiently within each block.
We compute the pairwise slew times between all fields assigned to a block, and then solve the Traveling Salesman Problem (TSP)
as an ILP problem in Gurobi\footnote{See \url{http://examples.gurobi.com/traveling-salesman-problem/} for a TSP solver implemented with Gurobi.}
using the cutting plane method \citep{Dantzig:54:TSP}.
In our application the quantity to be minimized is not the total length of the salesman's tour, but the total time spent slewing between fields within the block.

Since the P48 is an equatorial telescope, the slew time between fields using the hour angle and declination axes of the telescope do not vary with time. 
However, slews of the dome are azimuthal and so must be computed for each time block individually.
Because the same field may be requested by multiple observing programs, we apply a penalty factor to prevent the same field from being observed multiple times consecutively within a block, reducing the redundancy of the repeated exposures.

\subsection{Re-solving within the Night} \label{sec:resolve}

It is computationally feasible to resolve the entire optimization problem repeatedly within the night to account for time lost to weather, TOOs, or other schedule disruptions.
However, once time is lost during the night there is a complex tradeoff in determining which observing sequences to complete.  
One option would be complete some observing sequences exactly as requested and omit others entirely.
Another possibility would be accept partial completion of the remaining request sets, but this may limit the scientific usefulness of the observations.
To avoid making program-dependent decisions, we implement recomputes in a more limited way: at each block boundary, the best un-observed requests from earlier in the evening are reassigned to any unused time in the current block.

\subsection{Comparison to other ILP Scheduling Algorithms} \label{sec:other_ilp}

Our ILP algorithm differs in important ways from those of LCO \citep{Lampoudi:15:LCOGTScheduler} and ALMA \citep{Solar:16:SchedulingModel}.
Because LCO and ALMA are scheduling scientifically disparate observations, both schedulers use the TAC-assigned priority to provide an overall objective function.
Beyond simple acceptability constraints, the schedulers do not weight by the relative quality of an observation at any given time.
In contrast, because ZTF is simply an imaging survey and all surveys have equal priority, we are free to optimize an objective function (\S \ref{sec:metric}) that explicitly and self-consistently accounts for the time-varying quality (and hence scientific value) of any given exposure.
Additionally, because the observations scheduled by LCO and ALMA are long relative to the time to transition between them, their scheduling algorithms do not attempt to account for these transitions.
For ZTF, readout and slew overheads account for about 25\% of any given exposure, and long slews and filter changes create even larger losses.  
Accordingly our approach sequences exposures within a block to explicitly minimize the time spend slewing, and our objective function penalizes filter changes for the time lost.

\subsection{Summary of Algorithm Features}

Our choice of this ILP algorithm was motivated by its strengths in handling cadenced observing within a night and in balancing several simultaneous surveys (\S \ref{sec:surveys}).
To our knowledge ZTF must attempt to execute more independent observing programs simultaneously than any other wide-field imaging surveys (typically five, in addition to TOO observations), so rigorous cadence control is required.
The complete night lookahead provided by our algorithm ensures that observations are scheduled for the best time in the night, accounting for the number of observations required, variations in airmass and sky brightness, and the competing demands of other surveys.
Our ILP constraints (Eqn. \ref{eqn:constraint_nreqs}) guarantee that the scheduler will provide the requested number of observations if a field is observed.
This capability is vital for the success of Observing Programs requiring many observations during the night.
For example, the ZTF Collaboration's Extragalactic High Cadence survey (\S \ref{sec:surveys}) requires six nightly observations per field in two filters over three or more hours, which would be challenging to schedule effectively without the lookahead provided by our algorithm.
The scheduler uses the past observing history to rigorously maintain night-to-night cadences and to enforce the time allocated to the various surveys.
The scheduler treats each survey uniformly and interleaves the requested observations.
The algorithm self-consistently trades the observing time lost to filter changes against potential improvements in the quality of the observations.
Finally, while these values are only a component of our scientifically-motivated optimization metric, we note that the scheduler is effective at observing near zenith and minimizing slew time (\S \ref{sec:performance}).

\subsection{Limitations}

Obtaining these characteristics required accepting some tradeoffs in the capabilities of the scheduler.
Notably, our algorithm does not enable exact cadences or filter sequences within a night.
For example, it is not possible to schedule a $g$-band observation followed 12--15 minutes later by an $i$-band observation.
Rather, a total number of observations per filter is guaranteed, each separated by roughly the time block size (here, 30 minutes).
Even that minimum separation is not strictly guaranteed, as the sequencing of the fields within a block is independent, and observations may occur near the end of one block and near the beginning of the next.
We do schedule a minority of surveys that require more controlled within-night cadences; we implement these as pre-defined queues that interrupt the operations of the primary scheduler (\S \ref{sec:implementation}).

While our Traveling Salesman solution (\S \ref{sec:tsp}) minimizes the slew time within a given block, the initial block assignment does not account for the slew time between the fields.  
Accordingly our algorithm cannot be said to globally minimize slew time, although in practice we find that slew overheads are small (\S \ref{sec:performance}).

It is possible that the scheduler does not assign enough observations to a specific block to fill it\footnote{This is known as ``slack'' in the optimization literature, and is also a feature of the LCO scheduler \citep{Lampoudi:15:LCOGTScheduler}.}.
This is because our constraint on the number of observations per block (Equation \ref{eqn:constraint_nperslot}) is less than or equal to the number of observations that would fill the block, not a strict equality.
Strict equality creates overconstrained models that cannot be solved.
In general, the scheduling algorithm is subject to the details of the input observing programs. 
If a large fraction of the observing time is concentrated on a narrow area of the sky, for instance, there is no way for the scheduler to manufacture unrequested observations to fill other parts of the night.
In practice, we manage this issue by simulating potential observing strategies in advance when possible.
Additionally, re-solves during the night (\S \ref{sec:resolve}) can fill in previously unscheduled time with scheduled observations that were missed.
Finally, we implement a ``fallback queue'' to ensure that useful observations can be obtained if the main queue runs empty.  
To date this fallback time has largely been used to improve sky coverage for reference image building.
Without re-solves, typically the amount of slack in the schedule is a few percent if the input observing programs are well-balanced.

Finally, our current scheduler implementation does not yet dynamically adapt to changing observing conditions within the night due to the additional operational complexity and potential for schedule thrashing. 
We do not attempt cloud avoidance, for instance, or adjust to changes in seeing.
Such extensions are possible. 
One approach would be to maintain the overall block structure but conduct more extensive re-optimization at the block boundaries, and use a more dynamic selection of the next target field within a block to handle short-timescale variations.

\section{Implementation} \label{sec:implementation}

We have implemented the scheduling algorithm as a Python library, which is publicly available\footnote{\url{https://github.com/ZwickyTransientFacility/ztf_sim}} under an open source license.
The scheduler code takes advantage of a range of open-source Python libraries, including \texttt{Astropy} \citep{2018AJ....156..123T}, \texttt{Astroplan} \citep{astroplan2018},
and \texttt{pandas} \citep{pandas:2010}.
We use the commercial optimization package \texttt{Gurobi}\footnote{\url{http://www.gurobi.com/}} \citep{gurobi} under an academic license to perform the core ILP optimization.
While some attempt has been made to make the scheduler interfaces telescope agnostic, the library does encode assumptions specific to the ZTF use case.

Our objective function (\S \ref{sec:metric}) requires a detailed sky brightness model. 
We trained a gradient boosted tree model as implemented in \texttt{xgboost}  \citep{Chen:2016:XST:2939672.2939785} on historical data from ZTF (and initially PTF). 
Our model predicts the sky brightness in each filter as a function of telescope pointing altitude and azimuth, sun altitude, and moon altitude, moon distance, and moon illumination fraction.

The scheduler library can be run both in simulation mode (using historical weather data from PTF) as well as in operations.
For on-sky scheduling, we use an \texttt{aiohttp}\footnote{\url{https://aiohttp.readthedocs.io/}} webserver on the primary host computer of the ZTF Robotic Observing System \citep[ROS;][]{tmp_Dekany:18:ZTFObservingSystem}.
The webserver calls the scheduling library and provides a RESTful interface for command and status information.

We run the optimizer for five minutes before the start of the night's observations using two cores of the host machine, which yields satisfactory results without interfering with other robotic operations.
Moving the scheduler to a dedicated host would enable us to obtain equivalent performance in a shorter time by parallelizing over a larger number of CPUs.
The Gurobi solver library offers native parallelization by initializing multiple candidate solutions on different threads and concurrently optimizing each, terminating when one thread obtains a solution.
The memory footprint during nightly optimization can be as high as 700\,MB, dropping to about 500\,MB in sustained operations, although we have not attempted to optimize these values.

The ROS system obtains the required evening and morning calibration observations; the scheduler is responsible for selecting on-sky science observations. 
The ROS software also updates focus through the night using telemetry from the 2k$\times$2k focus CCDs on the perimeter of the mosaic.
Focus observations and updates occur concurrently with science observations and do not create additional overheads.

While most ($\gtrsim$90\%) of ZTF's Observing Programs are scheduled using our ILP algorithm, a subset require precise sequencing over continuous time blocks.  
These programs we implement as simple ``list queues'' that prescribe an expected start and stop time and a defined sequence of exposures to take.
We use this mechanism both for pre-planned observations as well as Target of Opportunity triggers.
A monitoring thread checks for the presence of such timed queues every ten seconds and switches to the appropriate queue if its validity window has started.
If list queues are planned before the start of the night's observing and will take at least one complete block, the ILP optimizer omits those blocks from scheduling the primary ZTF programs.

\section{ZTF Surveys} \label{sec:surveys}

ZTF observing time is divided between three major programs: 
public surveys facilitated by an award from the NSF Mid-Scale Innovations Program (MSIP; 40\% of the telescope time);
surveys designed by the members of the ZTF Collaboration (40\%);
and surveys selected each semester by the Caltech TAC (20\%).
The ZTF scheduler attempts to achieve this balance each calendar month, roughly the interval in which collaboration and Caltech sub-programs change.
The scientific goals of the surveys are discussed in \citet{tmp_Graham:18:ZTFScience}.

All surveys select fields from a discrete field grid\footnote{See \url{https://github.com/ZwickyTransientFacility/ztf_information}.}.
The current surveys all use the ``primary'' grid, which covers the entire sky with an average overlap between fields of about  0.29$^\circ$ in RA and
0.26$^\circ$ in Dec.
The average spacing between fields in the primary grid is 7.2$^\circ$ North--South and 7.0$^\circ$ East--West.
Upgrades to the P48 drive motors enables slews between adjacent fields within the 8.3\,sec CCD readout time.
The primary grid is arranged to align with $b=0^\circ$ of the Galactic Plane to improve the efficiency of the MSIP surveys.  
It also ensures good coverage of nearby galaxies (M31, M33, M51, M101, etc.).
A secondary grid, offset from the first by roughly half a field in RA and Dec, fills in missing sky coverage due to the gaps between CCDs and provides additional depth for sky areas covered by vignetted corners of the focal plane in the primary grid.
The primary grid alone covers 87.5\% of the sky; with the addition of the secondary grid, spatial coverage increases to 99.2\%.

Table \ref{tab:surveys} provides a high-level overview of the major public and collaboration surveys.

\begin{table*}
\begin{center}
\caption{Major ZTF surveys, Year One.  
Partnership surveys transition on the fifteenth of the month. 
The High-Cadence Plane Survey substituted for the Extragalactic High Cadence survey for two weeks  in August 2018.
\label{tab:surveys}}
\begin{tabular}{llllll}
\hline
Survey & Total Survey Footprint & Inter-night Cadence & Nightly Cadence & Average Nightly Area & Time Allocated \\
\hline
\multicolumn{6}{c}{Public Surveys} \\
\hline
Northern Sky Survey & 23675 deg$^2$ & 3 days & 1 $g$, 1 $r$ & 4325 deg$^2$ & $40\% \times 85\%$ \\
Galactic Plane Survey & 2800 deg$^2$  & 1 day & 1 $g$, 1 $r$ & 1475 deg$^2$ & $40\% \times 15\%$ \\
\hline
\multicolumn{6}{c}{ZTF Collaboration Surveys (Year One)} \\
\hline
Extragalactic High Cadence & 3000 deg$^2$ & 1 day & 3 $g$, 3 $r$ & 1725 deg$^2$ & $40\% \times 67.5\%$, \\
& & & & & Mar.--Nov.\\
$i$-band & 10725 deg$^2$ & 4 day & 1 $i$  & 1975 deg$^2$ & $40\% \times 22.5\%$, \\
& & & & & Mar.--Nov.\\
Target of Opportunity & varies & varies & varies & varies & $40\% \times 10\%$ \\
High-Cadence Plane Survey & $\sim$2100 deg$^2$ & N/A & $\gtrsim$2.5\,hr continuous, $r$ & 95\,deg$^2$ & $40\% \times 80\%$, \\
& & & & & Aug., Nov.--Jan. \\
Twilight Survey & N/A & N/A & 4 $r$ & 425\,deg$^2$ & 12$^\circ$--18$^\circ$ twilight, \\
& & & & & Nov.--Feb. \\
Asteroid Rotation Period   & N/A & N/A & $>25$ $r$ & 600 deg$^2$ & $40\% \times 80\%$, \\
& & & & & Jan.--Feb. \\
\hline
\end{tabular}
\end{center}
\end{table*}

\subsection{Public Surveys}

The ZTF public surveys were defined in the ZTF proposal to the NSF MSIP program.
A ``Northern Sky Survey'' covers all fields with centers $\delta \geq -31^\circ$ and $|b| > 7^\circ$.
When a field is up, on every third night it is observed once in $g$-band and once in $r$-band, with a spacing of at least 30 minutes between observations to discriminate between transients and moving objects \citep[cf.][]{Miller:17:ColorMeIntrigued}.
The Northern Sky Survey is allocated 85\% of the public time (34\% of telescope time).

ZTF also conducts a Galactic Plane Survey using the remaining 15\% of the public time (6\% of telescope time).
Fields with  $\delta \geq -31^\circ$ and $|b| \leq 7^\circ$ are visited twice each night they are visible, with one observation in $g$-band and one in $r$-band, again separated by at least 30 minutes.

\begin{figure}
\includegraphics[width=\columnwidth]{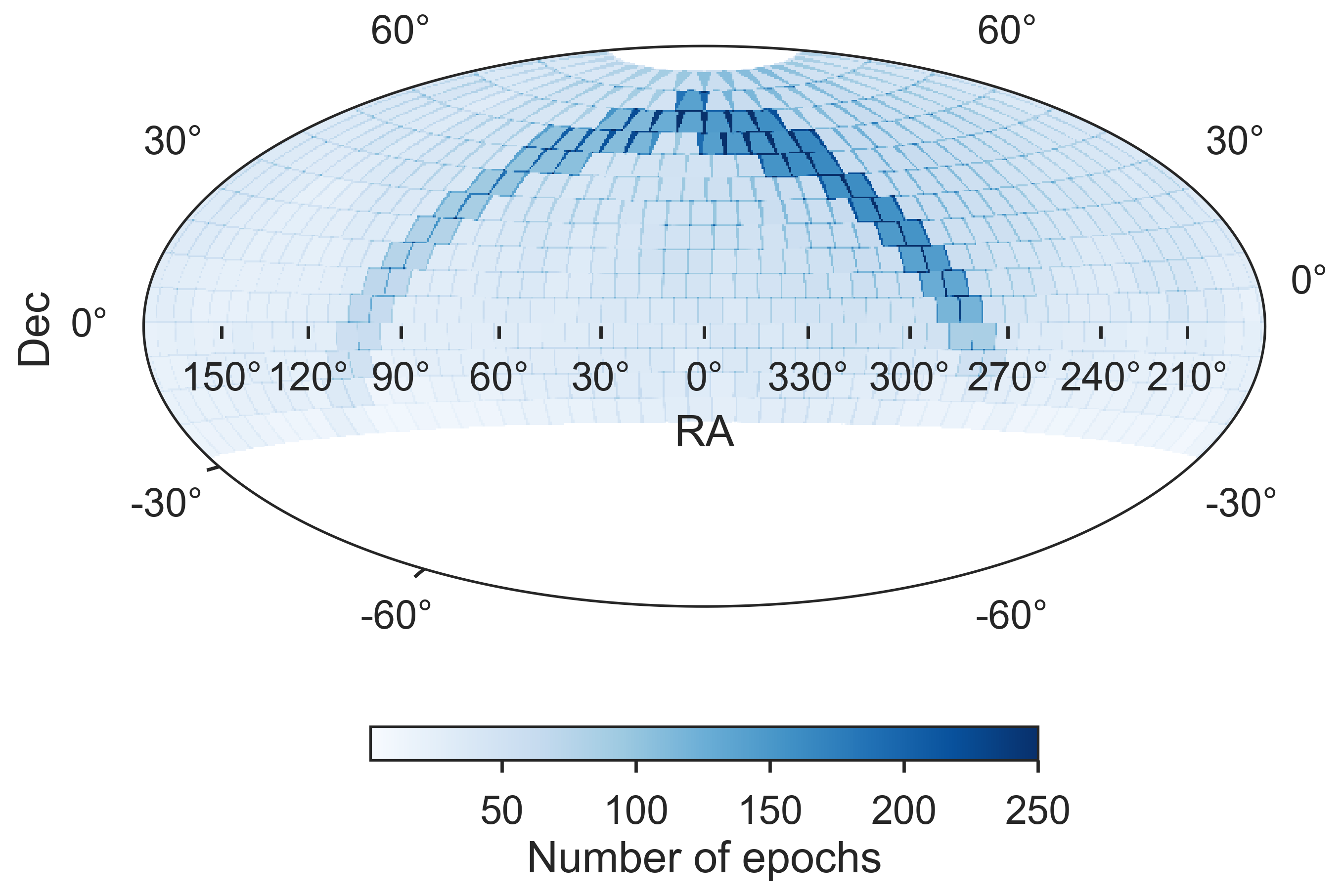}
\includegraphics[width=\columnwidth]{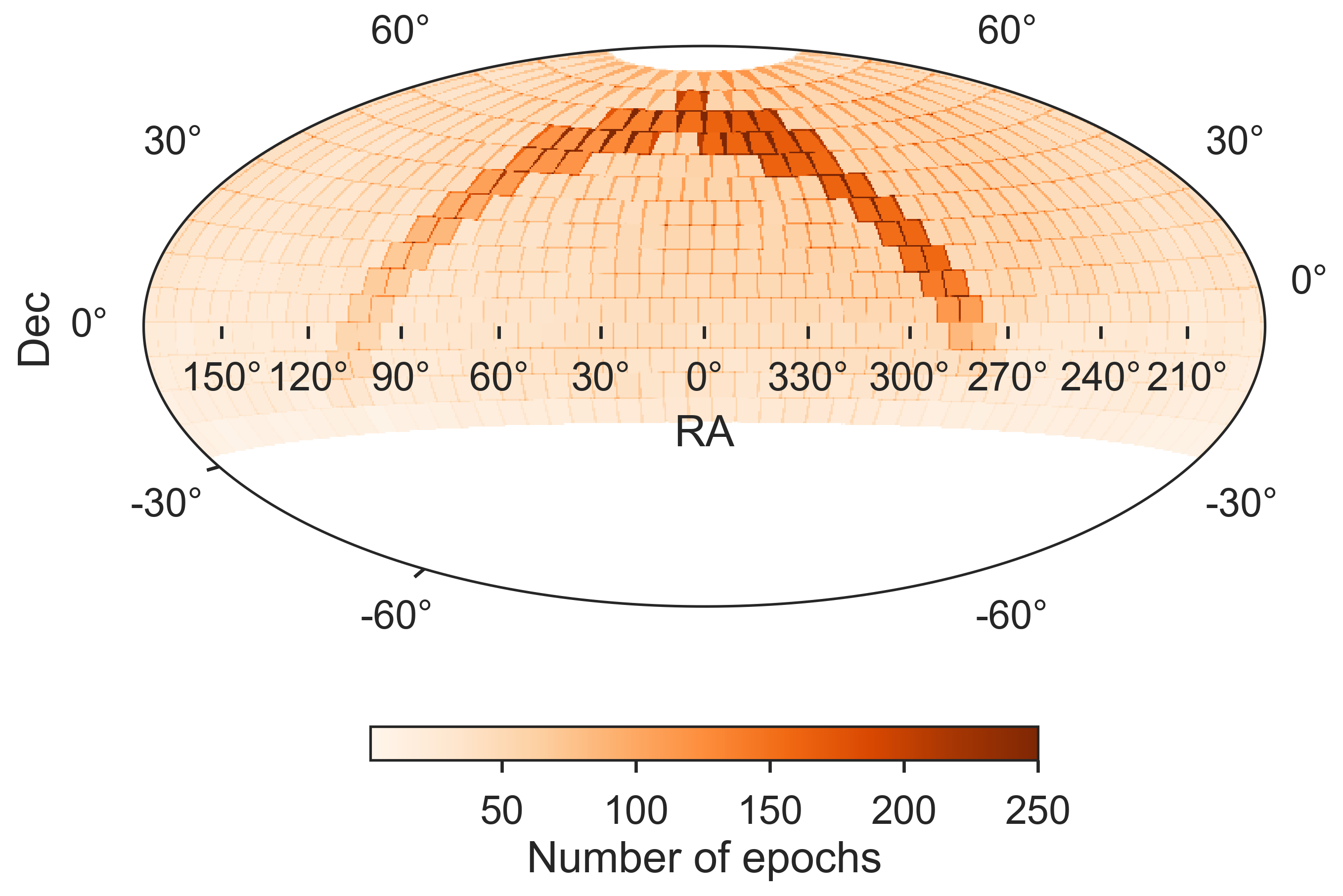}
\caption{Number of epochs obtained by the MSIP surveys across the sky in $g$-band (blue, top) and $r$-band (orange, bottom) to date.
\label{fig:msip_coverage}}
\end{figure}

Figure \ref{fig:msip_coverage} shows the $g$ and $r$ band sky coverage of the MSIP surveys to date.

Public alerts are issued in near real-time for all sources identified in image differencing from the public surveys \citep{tmp_Masci:18:ZTFDS, tmp_Patterson:18:ZTFAlertDistribution}.
Additionally, images, catalogs, and direct imaging lightcurves \citep{tmp_Masci:18:ZTFDS} will be released in data releases beginning in 2019.

We plan to continue these surveys in their present form through the first half of the three-year ZTF survey.
At that point we will assess the scientific returns from ZTF and the broader time-domain landscape and evolve the public surveys accordingly\footnote{The ZTF MSIP PI will select the public surveys in consultation with the ZTF Community Science Advisory Committee.}.

\subsection{ZTF Collaboration Surveys}

The ZTF Collaboration defined an initial slate of surveys for the first year of ZTF operations, although an extended commissioning period meant that the total time the surveys were executed is about 11 months.
Five major surveys were approved, with approximately two collaboration surveys plus Target of Opportunity observations active at any one time.

The bulk of the time (mid-March to mid-November) was dedicated to two extragalactic surveys: a high-cadence survey of 6 visits nightly (3 in $g$-band and 3 in $r$-band; 67.5\% of the collaboration time, or 27\% of the total time) optimized for the discovery of young supernovae and other fast transients, and a slow, wide $i$-band survey (one visit per field every four nights; 22.5\% of the collaboration time, or 9\% of the total time) designed to improve the cosmological constraining power of ZTF Type Ia supernovae.
In the future, co-adding multiple images taken by the high-cadence survey within a night can provide additional sensitivity to faint transients as well as strongly-lensed supernovae \citep{Goldstein:18:LensedSNe}.

Additionally, 10\% of the collaboration time (4\% of the total time) was reserved for Target of Opportunity observations of gamma-ray bursts, neutrino counterparts, gravitational wave triggers from LIGO and VIRGO, and Near-Earth Objects.

Two weeks in August and two months from mid-November to mid-January were allocated to very high cadence observations of Galactic Plane fields. 
A typical observation pattern was to alternate between two adjacent fields continuously for 2.5\,hours on two consecutive nights in $r$-band.
These ``continuous cadence'' observations enabled more sensitive searches for short-period binaries and stellar outbursts.

For three months from mid-November to mid-February, the period from 12 degree to 18 degree evening and morning twilight was devoted to the search for Near-Earth Objects at small solar elongation, with four visits over a ~30 minute period in $r$-band separated by 5-10 minutes.

Finally, during the period from mid-January to mid-February, a high-cadence survey near opposition will obtain tens of nightly observations per field in order to identify fast-rotating asteroids.

\begin{figure}
\includegraphics[width=\columnwidth]{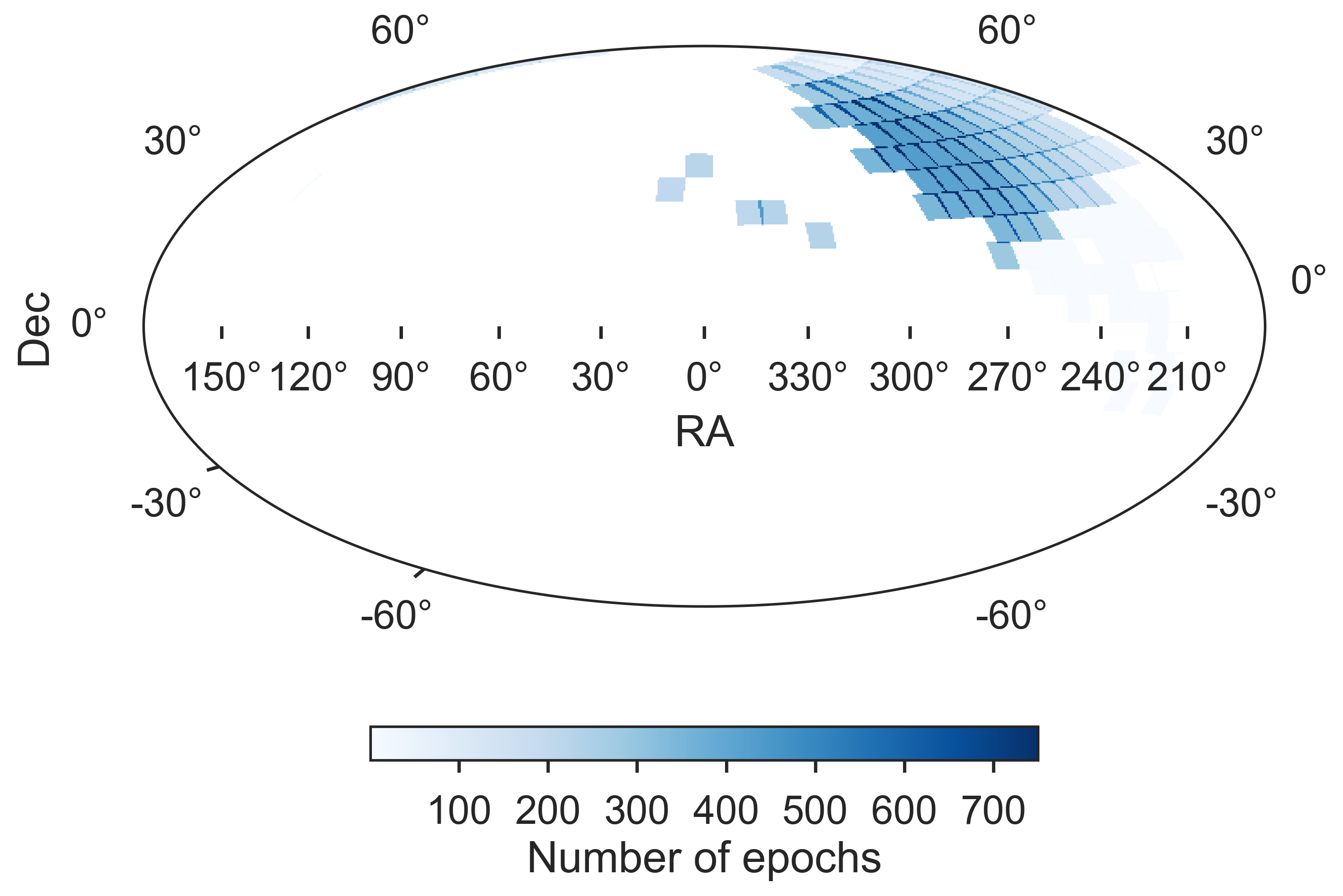}
\includegraphics[width=\columnwidth]{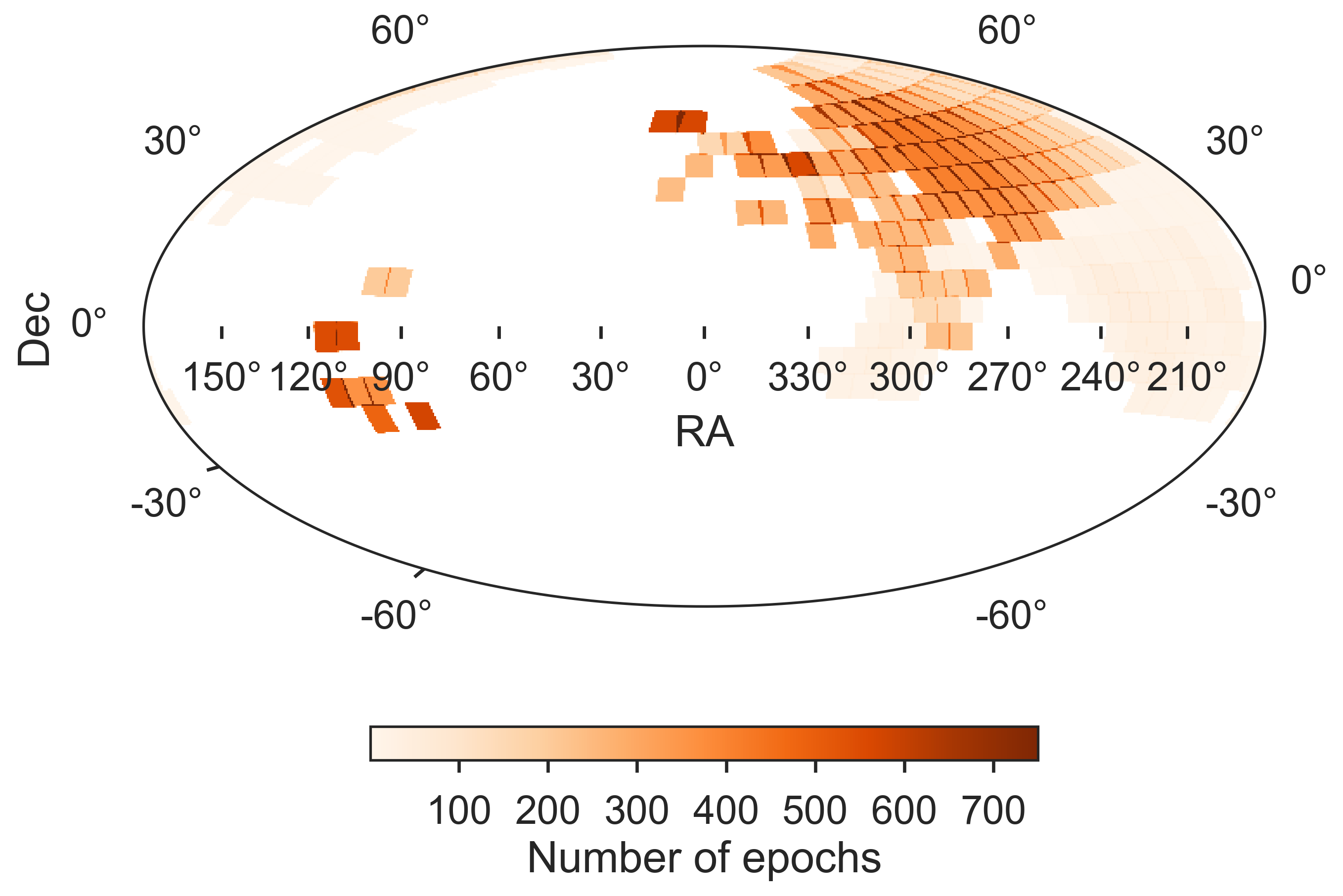}
\includegraphics[width=\columnwidth]{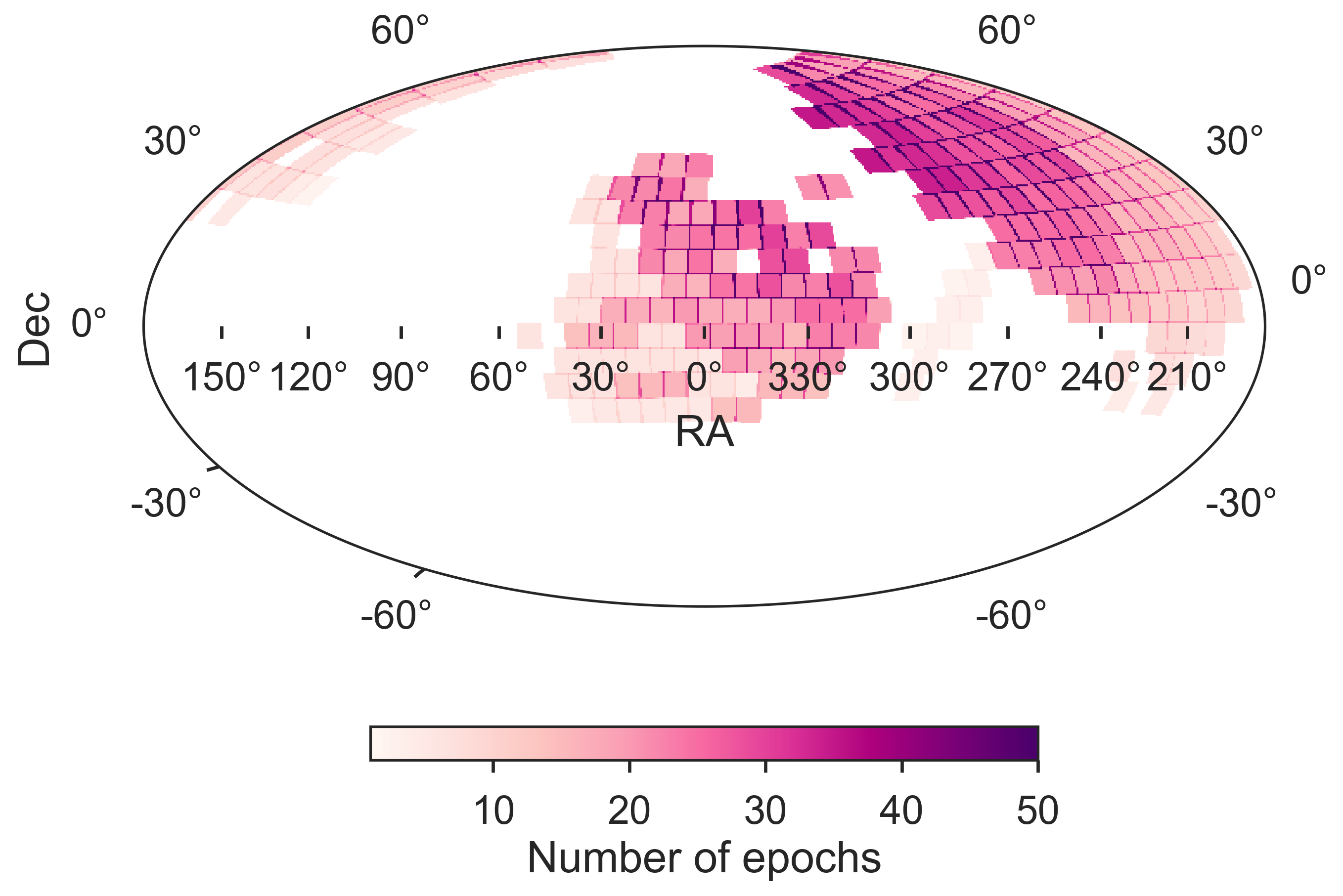}
\caption{Number of epochs obtained by the ZTF Collaboration surveys across the sky in $g$-band (blue, top), $r$-band (orange, middle), and $i$-band (pink, bottom) to date.
$g$-band observations are almost exclusively in the Extragalactic High-Cadence fields, and $i$-band observations in the $i$-band survey fields.
Several surveys contribute to the $r$-band coverage, including the Extragalactic High Cadence survey, the High-Cadence Plane Survey, and the Twilight Survey.
\label{fig:collaboration_coverage}}
\end{figure}

Figure \ref{fig:collaboration_coverage} shows the $g$ and $r$ band sky coverage of the collaboration surveys to date.

New ZTF Collaboration surveys will be selected for observations in 2019.

Images, catalogs, and lightcurves for data obtained during collaboration surveys will be released publicly during scheduled data releases after an 18 month proprietary period.

\subsection{Caltech Surveys}

Surveys selected by the Caltech TAC have included programs optimized for the discovery of transient, variable, and moving objects, with particular priority given to cadences and sky areas not being surveyed by the collaboration.
As these surveys are proposed and led by individuals we do not detail them further in this manuscript.
Data releases for these surveys are the responsibility of the proposer.

\section{Performance} \label{sec:performance}

\subsection{Simulated Performance}

To compare the performance of our ILP algorithm to a simple greedy optimizer, we simulated the May 2018 observing programs using both optimizers with realistic weather losses.
Both algorithms attempted to maximize our survey speed metric (Eqn. \ref{eqn:V}): the ILP algorithm optimized the form of the objective function in Eqn. \ref{eqn:objective}, while the greedy algorithm optimized the instantaneous volumetric survey speed $\dot{V}$ (Eqn. \ref{eqn:Vdot}).
Both approaches yielded comparable numbers of exposures per hour.
However, the ILP approach provided an 9\% increase in the metric (Eqn. \ref{eqn:V}) summed over all exposures.
It also scheduled observations closer to zenith, with a median airmass of 1.11 compared to 1.20 for the greedy approach.
Perhaps surprisingly, the greedy scheduler yielded fewer filter changes, averaging 2.0 per night compared to 3.6 per night.
Additionally, the ILP solution produced 4\% slack before within-night re-optimization.

The importance of the lookahead provided by the ILP algorithm is most clearly demonstrated by the sequence completion fraction---the fraction of observed fields for which the scheduler obtains all of the desired nightly observations.
Including the effects of weather losses, the greedy algorithm completed an average of 64\% of the MSIP Northern Sky Survey observations, 79\% of the MSIP Galactic Plane Survey observations, and 72\% of the collaboration Extragalactic High-Cadence observations.
In contrast, the ILP scheduler completed 81\% of the requested observations for each of the same surveys.

\subsection{On-Sky Performance}

The scheduler has performed effectively during on-sky operations.
It has scheduled more than 120,000 observations since the start of formal survey operations.
Overall balance between the MSIP, ZTF Collaboration, and Caltech observing programs was maintained, with 42\% of scheduled observations conducted in the MSIP surveys, 40.2\% in the collaboration surveys, and 18.8\% in the Caltech surveys.
The slight shortfall in the Caltech programs can be attributed in part to short intervals when no Caltech programs were available or they did not fill the entire time allocation.

\begin{figure}
\includegraphics[width=\columnwidth]{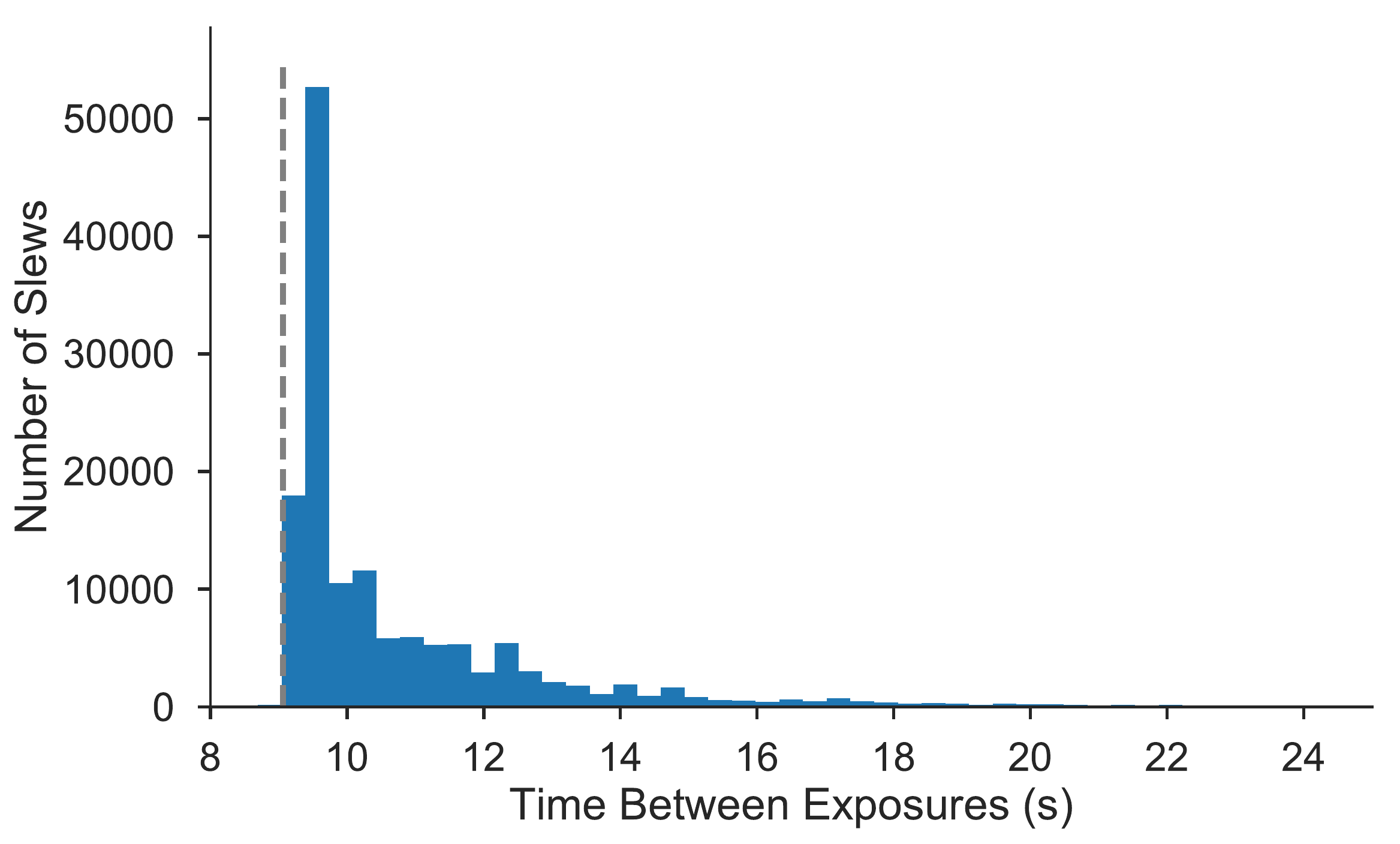}
\caption{Histogram of time elapsed between the end of one observation and the start of the next. 
The vertical dashed line indicates the shortest possible time between exposures ($\sim$9.1\,sec) due to readout time and the shutter opening and closing.
\label{fig:overhead_time_hist}}
\end{figure}

\begin{figure}
\includegraphics[width=\columnwidth]{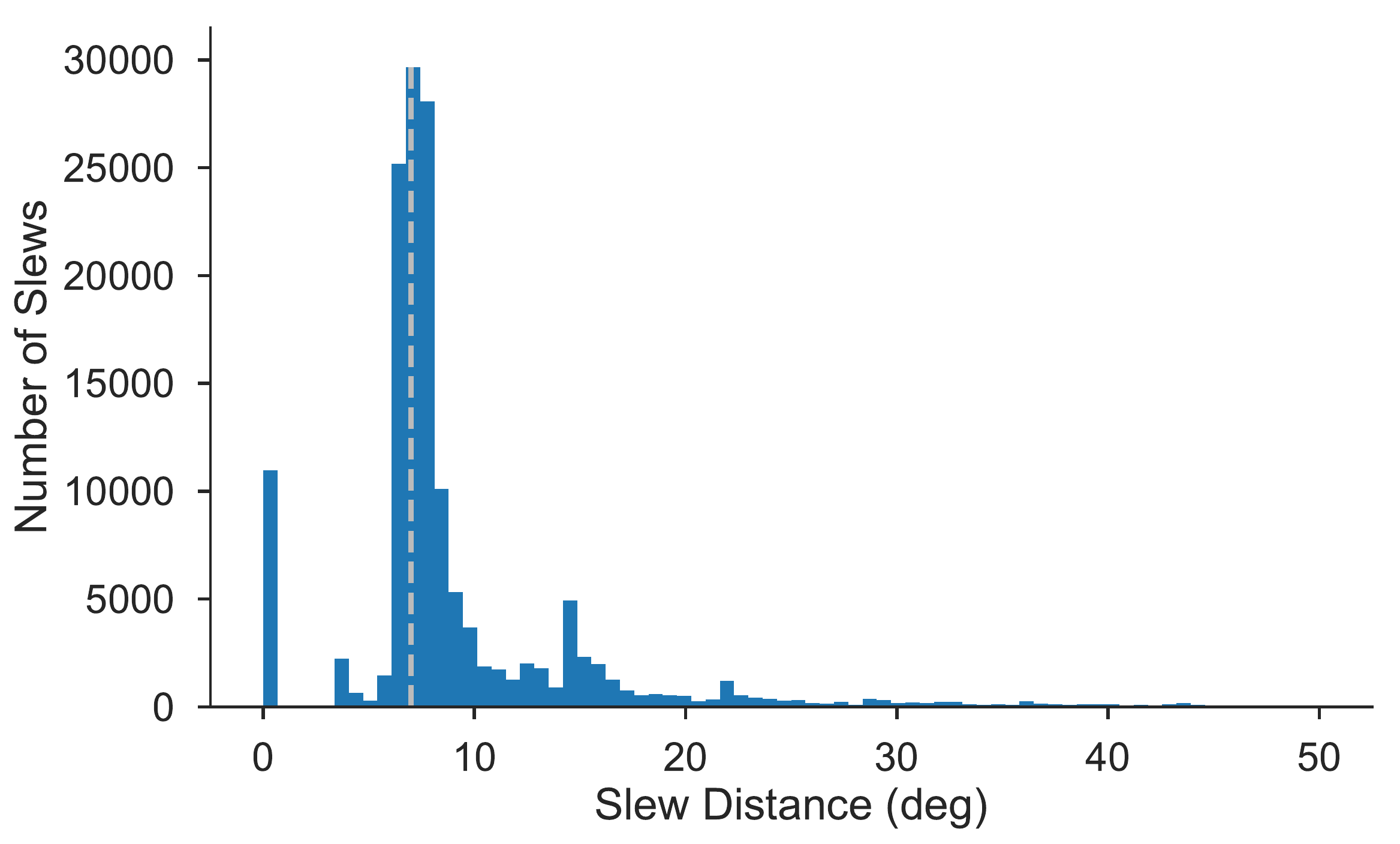}
\caption{Histogram of total distance slewed between observations.  The vertical dashed line at 7$^\circ$ indicates the average East--West distance between two adjacent fields in the same grid, although the exact grid spacing varies slightly with declination.
\label{fig:slew_distance_hist}}
\end{figure}

\begin{figure}
\includegraphics[width=\columnwidth]{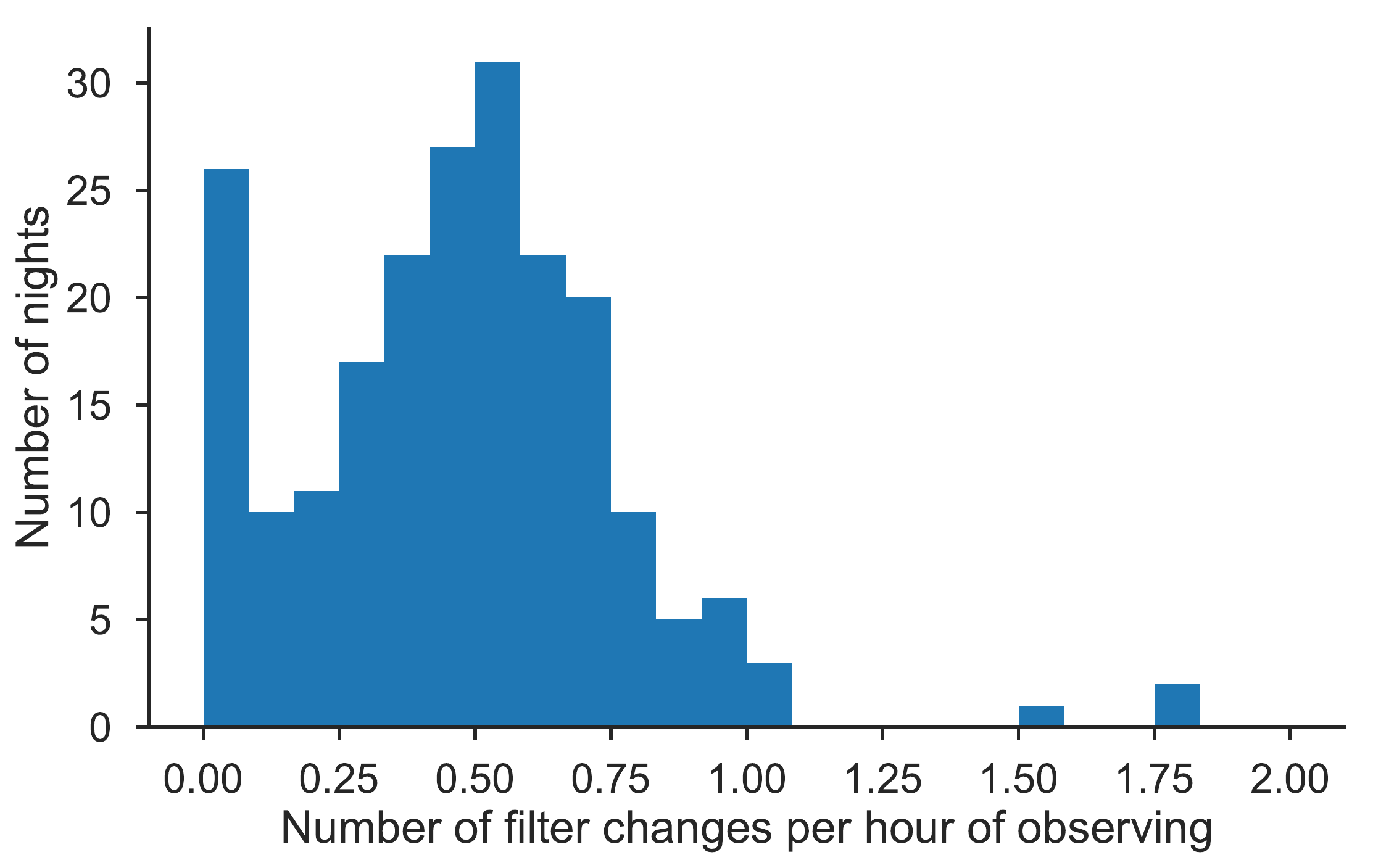}
\caption{Histogram of the number of filter exchanges per hour, computed on a nightly basis.
Nights shortened by weather may have no filter exchanges and hence appear as zero exchanges per hour.
\label{fig:filterchange_hist}}
\end{figure}

\begin{figure}
\includegraphics[width=\columnwidth]{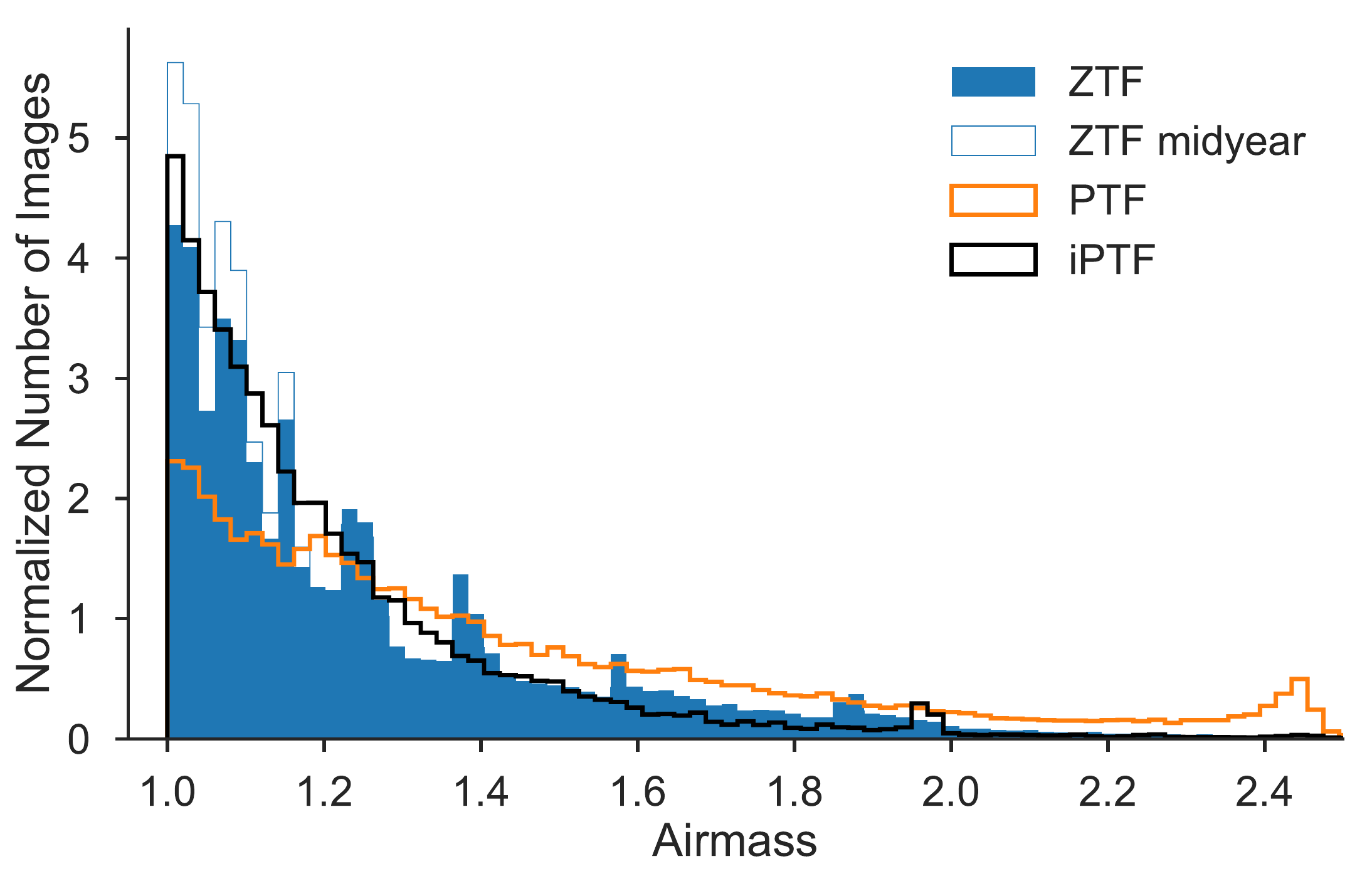}
\caption{Histogram of airmass values for ZTF (filled blue), PTF (black), and iPTF (orange).
The structured peaks in the ZTF histogram are due to the wider spacing of the fields compared to PTF.
During the late spring to early fall ZTF observed at lower airmass (light blue histogram) due to the distribution of collaboration fields and the shorter nights.
\label{fig:airmass_hist}}
\end{figure}

\begin{figure}
\includegraphics[width=\columnwidth]{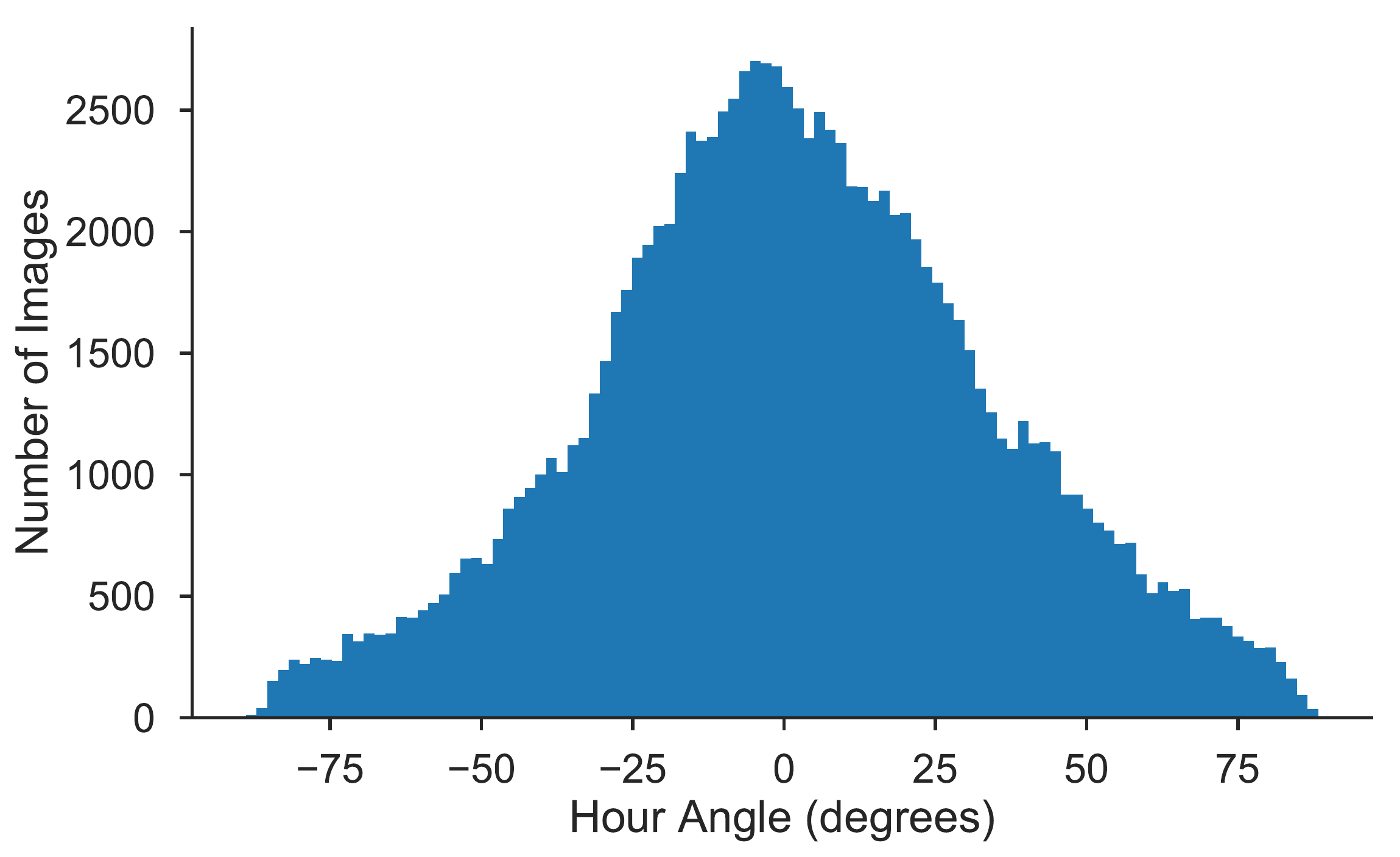}
\caption{Histogram of hour angle values for ZTF. 
\label{fig:hour_angle_hist}}
\end{figure}

\begin{figure}
\includegraphics[width=\columnwidth]{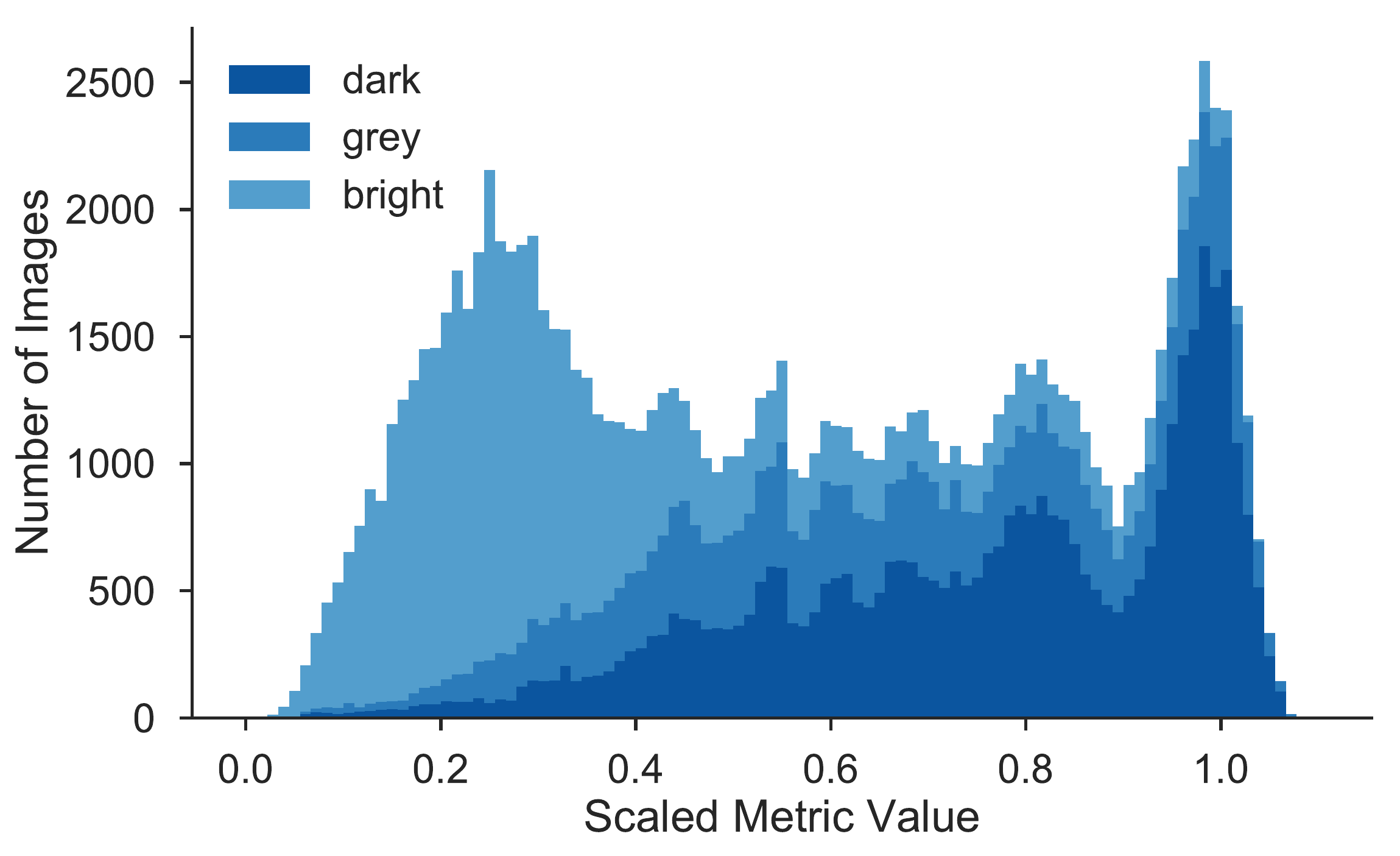}
\caption{Histogram of the metric values per image.
Colors indicate tertiles of moon phase, with dark blue corresponding to dark time (0--33\% moon phase), medium blue indicating grey time (33--66\%), and light blue bright time (66--100\%).
Smaller scale structure is due to the discrete spacing of the ZTF field grid: some fields transit at higher airmasses depending on their declination.
\label{fig:metric_hist}}
\end{figure}

The scheduler uses the telescope efficiently, with the median time between observations of 9.9\,sec (Figure \ref{fig:overhead_time_hist}) and most slews of one field offset (Figure \ref{fig:slew_distance_hist}).
The tenth--90th percentile overhead times and slew distances are 9.4\,sec--14.9\,sec and 6.0$^\circ$--16.1$^\circ$ respectively.
Repeated exposures of the same field without slews have a median time between exposures of 9.4\,sec.
Filter exchanges occur less than once per hour during the vast majority of nights (Figure \ref{fig:filterchange_hist}).
While the ZTF filter exchanger hardware is designed to support a higher rate of filter changes, our penalty factor (Eqn. \ref{eqn:objective}) self-consistently trades the need for filter changes against the time lost during the exchange and prevents filter changes from occurring on every block boundary.
The choice of optimization metric leads observations to be preferentially scheduled around zenith (Figures \ref{fig:airmass_hist} and \ref{fig:hour_angle_hist}).
Figure \ref{fig:metric_hist} shows the resulting metric values, which vary sharply with moon phase.

\begin{figure}
\includegraphics[width=\columnwidth]{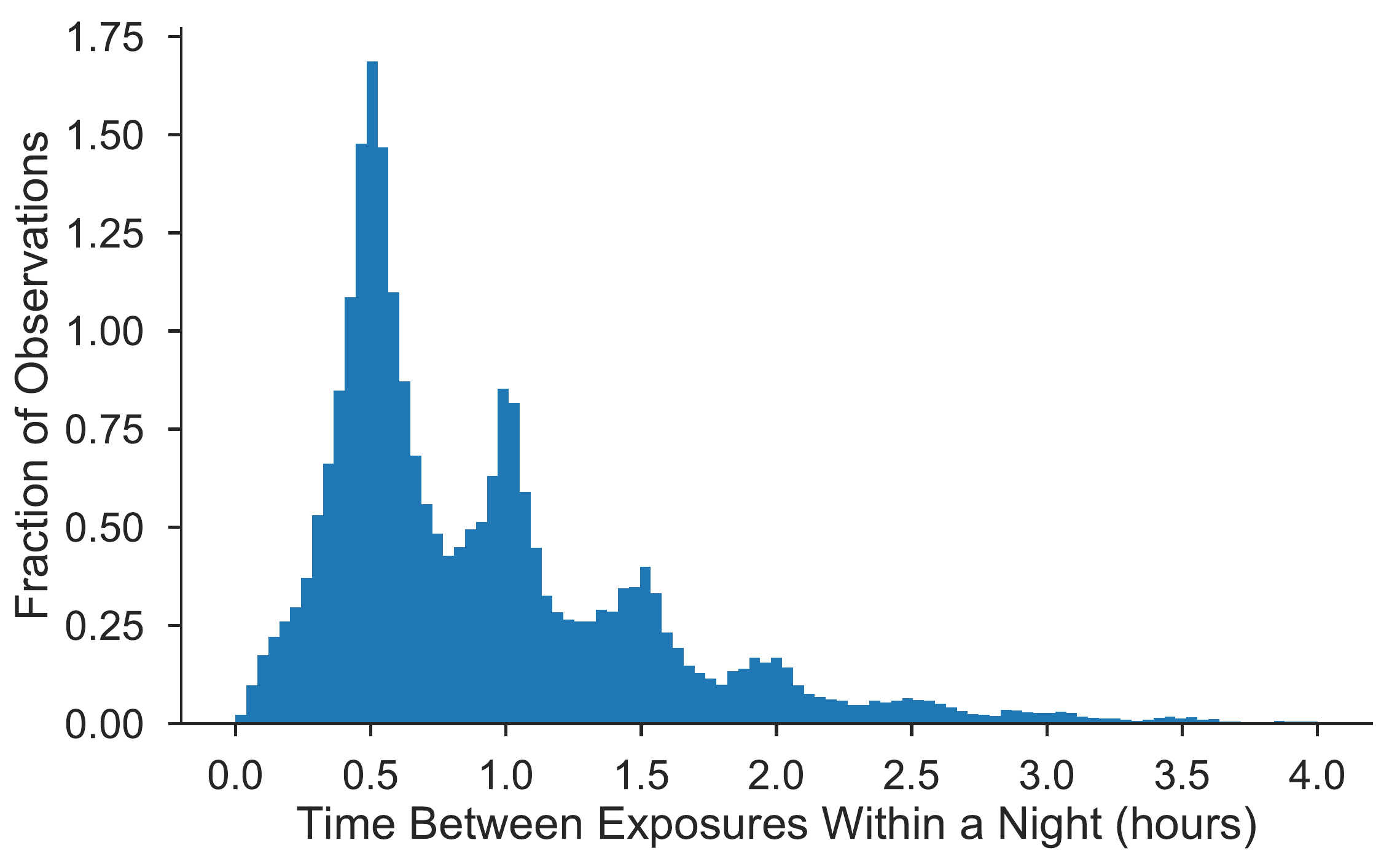}
\caption{Histogram of times between successive observations of a field by a given program within a night.
\label{fig:internight_cadence_hist}}
\end{figure}

\begin{figure*}
\includegraphics[width=\columnwidth]{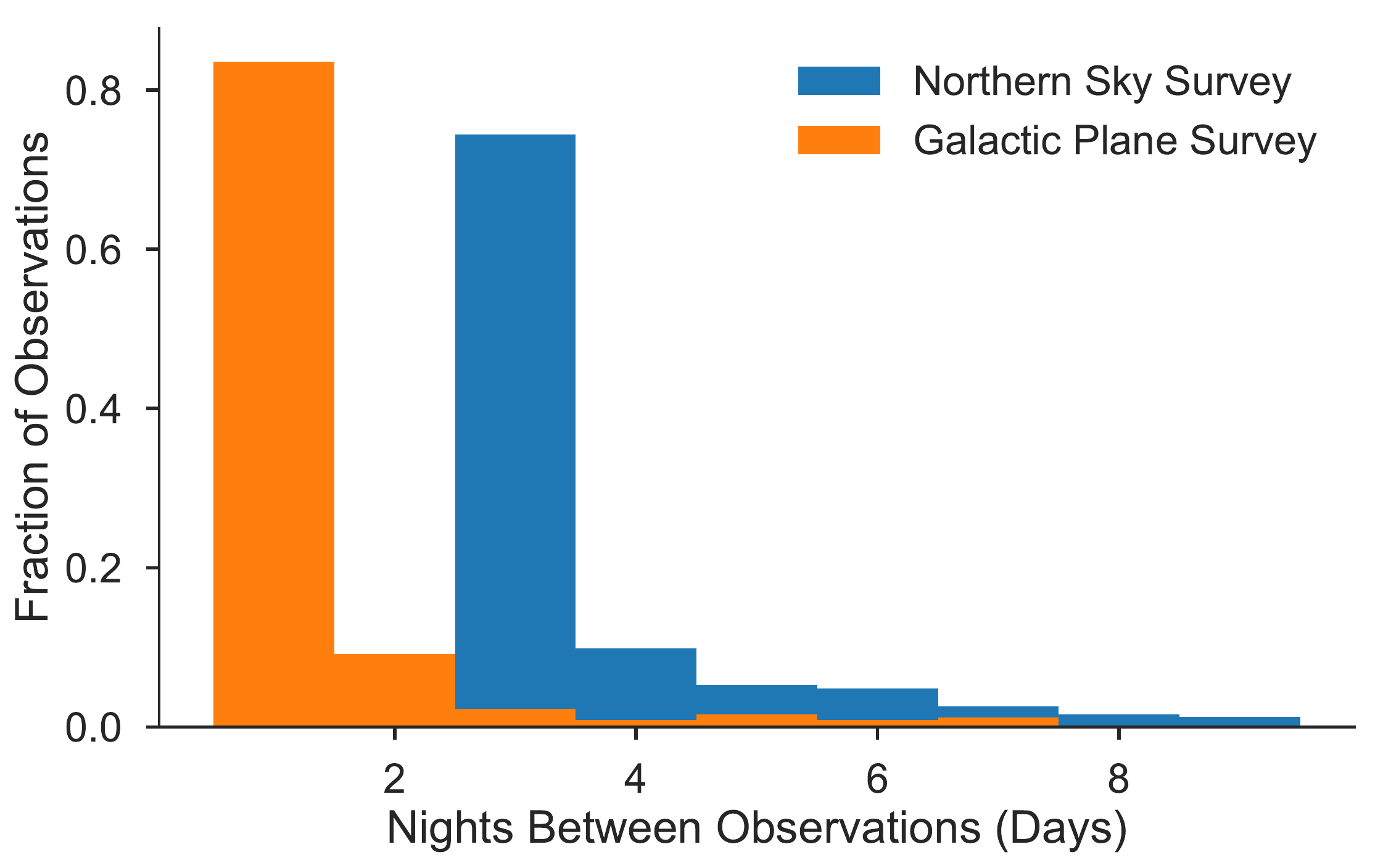}
\includegraphics[width=\columnwidth]{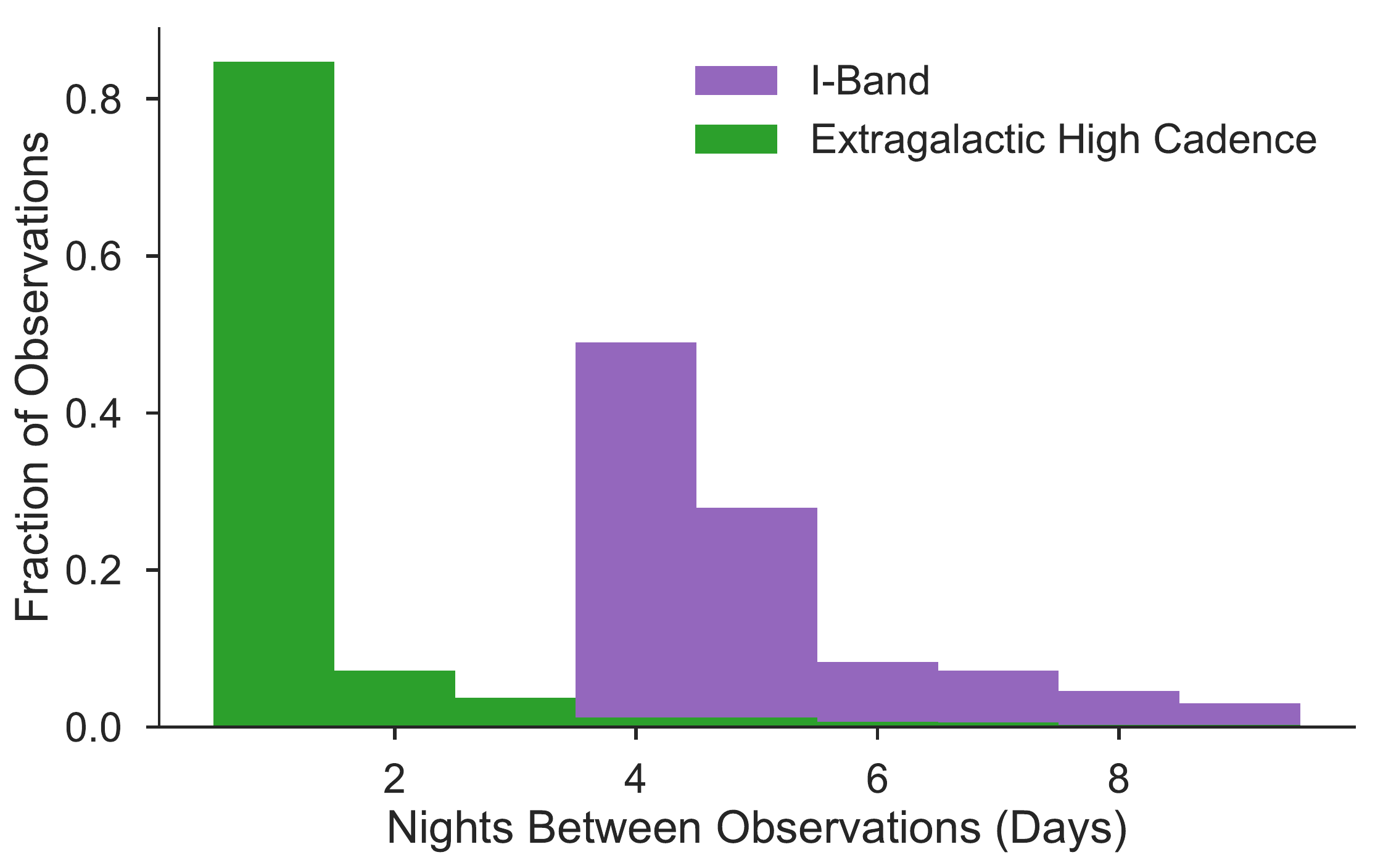}
\caption{Histogram of times between successive observations of a field by a given program from night to night.  Left: MSIP Northern Sky Survey (3-day cadence) and Galactic Plane Survey (1-day cadence).  Right: Partnership $i$-Band Survey (4-night cadence) and Extragalactic High Cadence Survey (1-day cadence).
Revisit times longer than the target cadence are due to weather and scheduling effects.
\label{fig:intranight_cadence_hist}}
\end{figure*}

The scheduler delivers the desired cadences.  
Eighty percent of all observations are spaced by at least 30\,min as desired for asteroid discrimination (Figure \ref{fig:internight_cadence_hist}).
The intra-night cadences for the major surveys are delivered as expected (Figure \ref{fig:intranight_cadence_hist}), with minimal tailing to longer-than-desired revisit times.

\begin{figure}
\includegraphics[width=\columnwidth]{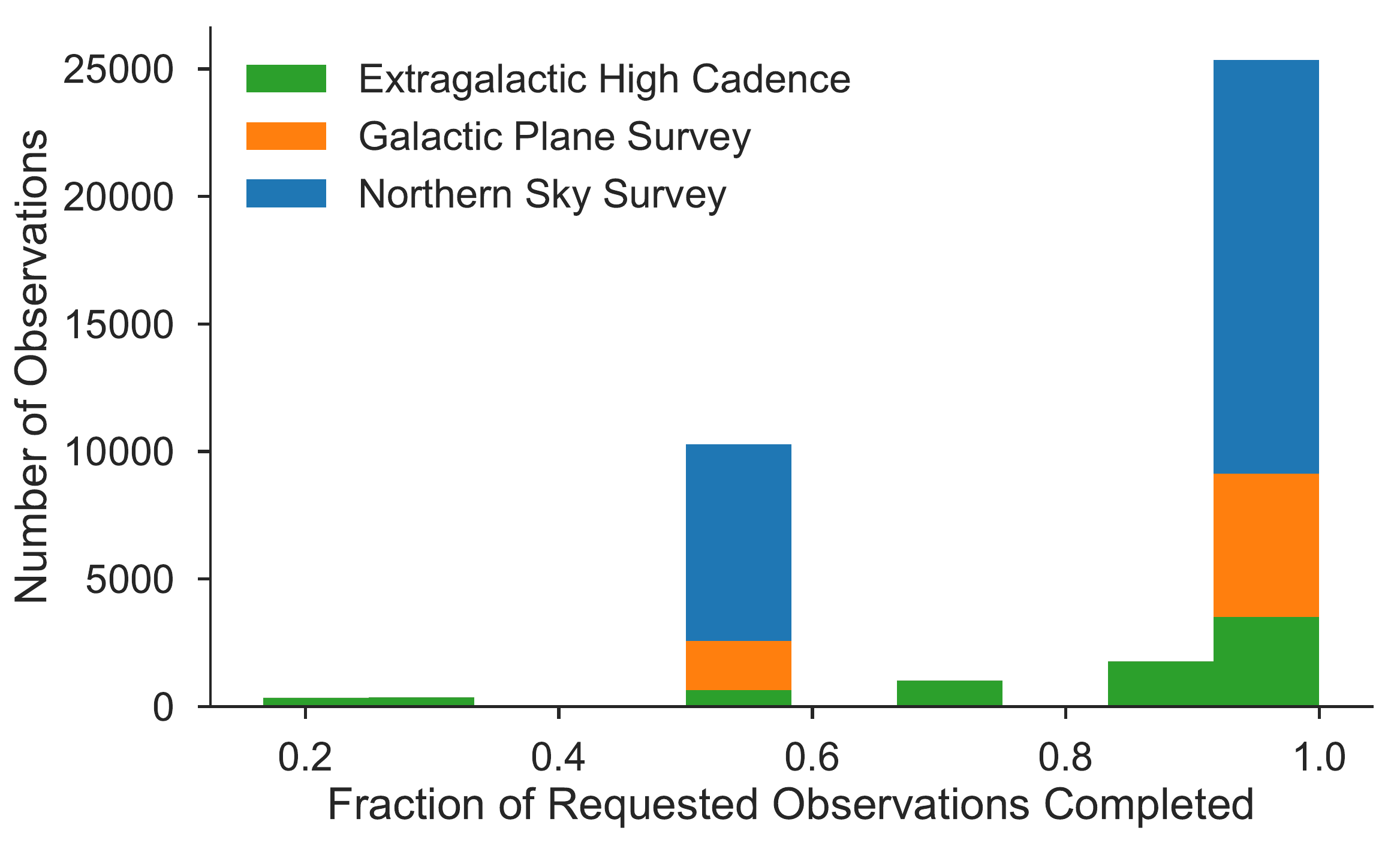}
\caption{Histogram of fractional sequence completion for the major ZTF surveys.
The MSIP Northern Sky Survey (All Sky) and Galactic Plane Survey (Nightly Plane) each request two observations per field, so have fractional completion of 0.5 or 1.0 on nights the field is observed.
The collaboration Extragalactic High Cadence Survey (High Cadence) has six observations nightly, so the fractional completion can range from 1/6--6/6.
\label{fig:sequence_completion}}
\end{figure}

Finally, the scheduler delivers a high fraction of completed observation sequences, averaging 84.6\% completion for the MSIP surveys and the collaboration high-cadence surveys (Figure \ref{fig:sequence_completion}).

\section{Conclusions} \label{sec:conclusions}

We have implemented a scheduling algorithm for wide-field imaging time-domain surveys that cleanly delineates three core concerns:

\begin{enumerate}
        \item The intrinsic quality of a specific image, as specified by signal-to-noise ratio or spatial volume probed (\S \ref{sec:metric}); this encapsulates image quality, sky background, airmass, and related terms.
        \item The scientific value of obtaining an image of a given field \textit{at a specific time}; these desired cadences are specified \textit{a priori}.
        \item The means of sequencing observations to maximize efficiency and throughput.
\end{enumerate}

A survey focused on a single class of astrophysical object could combine the first two goals, trading off the information gained from a high SNR observation now versus a low SNR observation later, using knowledge of lightcurve shape, periodicity, etc.
However, this combination is not possible for a general-purpose, wide-field survey.
Similarly, long-term planning could account for the uncertain availability of future observations (due to weather, instrument failures, etc.)\footnote{i.e., ``Expected Future-Discounted Information Gain'' \citep{Hogg:EFDIG}.}.

We suggest that this formalism would provide useful clarity to the problem of scheduling observations for LSST.
In particular, we argue that an appropriate scheduler for LSST would attempt to maximize the contribution of a night's observing to the total coadded depth of the survey, subject to the desired cadence constraints.
This is simply the approach developed here with a slightly modified objective function (\S \ref{sec:metric}). 
It directly optimizes the metric of interest without requiring intermediary features which intermix concerns of image quality, cadence, and efficiency \citep[cf.][]{Naghib:18:FeatureBasedScheduler}.
Since the number of exposures scheduled nightly for ZTF and LSST are comparable, our on-sky implementation demonstrates directly that this algorithm could be feasibly applied to LSST.
Further work would be needed to adapt our algorithm to meet all LSST requirements and rigorously compare its performance to other scheduling approaches, however.

The coming decade will see new surveys of unprecedented scale---imaging and spectroscopy, on the ground and in space.
To fully reap the scientific value of these large investments, astronomers must give sustained attention to the scheduling problems unique to each survey.
Cross-fertilization with research in the field of Operations Research may be of particular value.
Different surveys will necessarily require different algorithms and metrics, 
but thanks to increasingly powerful computing resources, new optimization approaches are now feasible.
Careful attention to scheduling can provide some of the most cost-effective improvements in science throughput available.

\acknowledgments
ECB thanks Eric Saunders, Ben Shappee, Lynne Jones, and Peter Yoachim for useful conversations.

 Based on observations obtained with the Samuel Oschin Telescope 48-inch and the 60-inch Telescope at the Palomar Observatory as part of the Zwicky Transient Facility project. ZTF is supported by the National Science Foundation under Grant No.\ AST-1440341 and a collaboration including Caltech, IPAC, the Weizmann Institute for Science, the Oskar Klein Center at Stockholm University, the University of Maryland, the University of Washington, Deutsches Elektronen-Synchrotron and Humboldt University, Los Alamos National Laboratories, the TANGO Program of the University System of Taiwan, the University of Wisconsin at Milwaukee, and Lawrence Berkeley National Laboratories. Operations are conducted by COO, IPAC, and UW. 

ECB is supported in part by the NSF AAG grant 1812779 and grant \#2018-0908 from the Heising-Simons Foundation.
ECB also acknowledges support from the Large Synoptic Survey Telescope, which is supported in part by the National Science Foundation through
Cooperative Agreement 1258333 managed by the Association of Universities for Research in Astronomy
(AURA), and the Department of Energy under Contract No.\ DE-AC02-76SF00515 with the SLAC National
Accelerator Laboratory. Additional LSST funding comes from private donations, grants to universities,
and in-kind support from LSSTC Institutional Members.
ECB is grateful for further support from the University of Washington College of Arts and Sciences, Department of Astronomy, and the DIRAC Institute. University of Washington's DIRAC Institute is supported through generous gifts from the Charles and Lisa Simonyi Fund for Arts and Sciences, and the Washington Research Foundation.

EOO is grateful for support by
grants from the  Willner Family Leadership Institute, Ilan Gluzman (Secaucus NJ), the Israel Science Foundation, Minerva, BSF, BSF-transformative, Weizmann-UK, and the I-Core program by the Israeli Committee for Planning and Budgeting and the Israel Science Foundation (ISF).

\facilities{PO:1.2m}

\software{
\texttt{Astropy} \citep{2018AJ....156..123T}, 
\texttt{Astroplan} \citep{astroplan2018},
\texttt{Numpy} \citep{numpy:2011}, 
\texttt{Scipy} \citep{scipy:2001}, 
\texttt{pandas} \citep{pandas:2010}, 
\texttt{Matplotlib} \citep{Hunter:2007},
\texttt{Seaborn} \citep{seaborn:2018},
\texttt{Scikit-Learn} \citep{scikit-learn:2011},
\texttt{xgboost} \citep{Chen:2016:XST:2939672.2939785},
\texttt{Gurobi} \citep{gurobi},
\texttt{makecite} \citep{makecite:2018}}

\bibliographystyle{aasjournal}
\bibliography{references}

\end{CJK*}
\end{document}